\documentclass[useAMS,usenatbib]{mnras}

\usepackage{pdflscape}
\usepackage{epstopdf}
\usepackage{graphicx}
\usepackage{textcomp}
\usepackage{courier}
\usepackage{caption}
\usepackage{amssymb,amsmath}
\usepackage{url}
\usepackage{deluxetable}
\usepackage{gensymb}

\usepackage{epsfig}\usepackage{epsf}
\usepackage{soul}

\title[Low-Polarization SN IIn 2014ab]{SN 2014ab: An Aspherical Type IIn Supernova with Low Polarization}
\author[Bilinski et. al.]{Christopher Bilinski$^{1}$\thanks{E-mail:
cgbilinsk@gmail.com}, Nathan Smith$^{1}$, G. Grant Williams$^{2}$, Paul Smith$^{1}$, 
\newauthor Jennifer Andrews$^{1}$, Kelsey I. Clubb$^{3}$,  WeiKang Zheng$^{3}$, Alexei V. Filippenko$^{3,4}$,
\newauthor Ori D. Fox$^{5}$, Griffin Hosseinzadeh$^{6}$, D. Andrew Howell$^{7,8}$, Patrick L. Kelly$^{9}$,  
\newauthor   Peter Milne$^{1}$, D. J. Sand$^{1}$, Jennifer L. Hoffman$^{10}$, Douglas C. Leonard$^{11}$,
\newauthor Samantha Cargill$^{3}$, Chadwick Casper$^{3}$, Goni Halevy$^{12}$, Haejung Kim$^{13}$,
\newauthor Sahana Kumar$^{3,14}$, Kenia Pina$^{3}$, Heechan Yuk$^{3,15}$\\
$^{1}$Steward Observatory, University of Arizona, 933 N. Cherry Avenue, Tucson AZ 85721, USA\\
$^{2}$MMT Observatory, Tucson, AZ 85721-0065, USA\\
$^{3}$Department of Astronomy, University of California, Berkeley, CA 94720-3411, USA\\
$^{4}$Miller Senior Fellow, Miller Institute for Basic Research in Science, University of California, Berkeley, CA  94720, USA\\
$^{5}$Space Telescope Science Institute, 3700 San Martin Drive, Baltimore, MD 21218, USA\\
$^{6}$Center for Astrophysics \textbar{} Harvard \& Smithsonian, 60 Garden Street, Cambridge, MA 02138-1516, USA\\
$^{7}$Department of Physics, University of California, Santa Barbara CA 93106-9530, USA\\
$^{8}$Las Cumbres Observatory, 6740 Cortona Drive Suite 102, Goleta, CA 93117-5575, USA\\
$^{9}$School of Physics and Astronomy, University of Minnesota, 116 Church Street S. E., Minneapolis, MN 55455, USA\\
$^{10}$Department of Physics \& Astronomy, University of Denver, 2112 East Wesley Avenue, Denver, CO 80208, USA\\
$^{11}$Department of Astronomy, San Diego State University, San Diego, CA 92812, USA\\
$^{12}$Department of Astrophysical Sciences, Princeton University, 4 Ivy Lane, Princeton, NJ, 08544, USA\\
$^{13}$Dr. Opinion, 201 Dolores Avenue, San Leandro, CA 94577, USA\\
$^{14}$Department of Physics, Florida State University, Tallahassee, FL 32306, USA\\
$^{15}$Homer L. Dodge Department of Physics and Astronomy, The University of Oklahoma, Norman, OK 73019, USA}

\begin{document}
\date{Accepted 0000. Received 0000; in original form 0000}

\pagerange{\pageref{firstpage}--\pageref{lastpage}} \pubyear{2020}

\maketitle 

\label{firstpage}

\begin{abstract}
We present photometry, spectra, and spectropolarimetry of supernova (SN) 2014ab, obtained through $\sim 200$ days after peak brightness.  SN~2014ab was a luminous Type IIn SN ($M_V < -19.14$\,mag) discovered after peak brightness near the nucleus of its host galaxy, VV~306c.  Prediscovery upper limits constrain the time of explosion to within 200 days prior to discovery.  While SN~2014ab declined by $\sim 1$\,mag over the course of our observations, the observed spectrum remained remarkably unchanged.  Spectra exhibit an asymmetric emission-line profile with a consistently stronger blueshifted component, suggesting the presence of dust or a lack of symmetry between the far side and near side of the SN.  The Pa$\beta$ emission line shows a profile very similar to that of H$\alpha$, implying that this stronger blueshifted component is caused either through obscuration by large dust grains, occultation by optically thick material, or a lack of symmetry between the far side and near side of the interaction region.  Despite these asymmetric line profiles, our spectropolarimetric data show that SN~2014ab has little detected polarization after accounting for the interstellar polarization.  This suggests that we are seeing emission from a photosphere that has only small deviation from circular symmetry face-on.  We are likely seeing a SN~IIn with nearly circular symmetry in the plane normal to our line of sight, but with either large-grain dust or significant asymmetry in the density of circumstellar material or SN ejecta along our line of sight. We suggest that SN~2014ab and SN~2010jl (as well as other SNe~IIn) may be similar events viewed from different directions.
\end{abstract}


\section{Introduction}
\label{sec:Int}

Type IIn supernovae (SNe~IIn) are observed when fast SN ejecta crash into dense circumstellar material (CSM).  Strong narrow and intermediate-width hydrogen emission lines indicate the presence of these CSM interaction regions \citep{1990MNRAS.244..269S,1997ARA&A,smith17}.  Because SNe~IIn are powered not only by emission from the SN ejecta, but also from CSM interaction, unraveling the geometry of the explosion can be complicated.  In order to better understand both the shape of the SN ejecta and the CSM, two different approaches (spectroscopy and spectropolarimetry) are commonly used.

Spectral line profile shapes can help clarify the geometry of the CSM or SN ejecta.  Blueshifted line profiles can arise when receding portions of the CSM or ejecta are extinguished by dust or occulted by the SN photosphere, while any type of line asymmetry might arise from an aspherical explosion or aspherical CSM.  The spectra of SNe~IIn often show signs of asphericity in the CSM and the SN ejecta revealed by line profiles (SN 1988Z: \citealt{1994MNRAS.268..173C}; SN 1995N: \citealt{2002ApJ...572..350F}; SN 1997eg: \citealt{2008ApJ...688.1186H}, SN 1998S: \citealt{2000ApJ...536..239L,2001ApJ...550.1030W,2005ApJ...622..991F,2012MNRAS.424.2659M}; SN 2005ip: \citealt{2009ApJ...695.1334S,2014ApJ...780..184K}; SN 2006jd: \citealt{2012ApJ...756..173S}; SN 2006tf: \citealt{2008ApJ...686..467S};  SN 2009ip: \citealt{2014MNRAS.442.1166M,2017arXiv170108885R}; SN 2010jl: \citealt{2012AJ....143...17S,2014ApJ...797..118F}; PTF11iqb: \citealt{2015MNRAS.449.1876S}; SN 2012ab: \citealt{2018MNRAS.475.1104B}; SN 2013L: \citealt{2017MNRAS.471.4047A}).  

One can also use spectropolarimetric data to measure the polarization and position angle of integrated light.  Although only a few SNe~IIn have spectropolarimetric data published (SN~1997eg: \citealt{2008ApJ...688.1186H}; SN~1998S: \citealt{2000ApJ...536..239L}; SN~2006tf: \citealt{2008ApJ...686..467S}; SN~2010jl: \citealt{2011A&A...527L...6P}; SN~2009ip: \citealt{2014MNRAS.442.1166M,2017arXiv170108885R}; SN~2012ab: \citealt{2018MNRAS.475.1104B}), every one observed so far has a high level of continuum polarization (1--3\%), seemingly indicating that strong asphericity is the standard for this class of objects. This conclusion may be biased if spectropolarimetric data of SNe~IIn with low or undetected polarization go unpublished, of course.

\begin{figure*}
\centering
\includegraphics[width=0.4\textwidth,clip=true,trim=0cm 0cm 0cm 0cm]{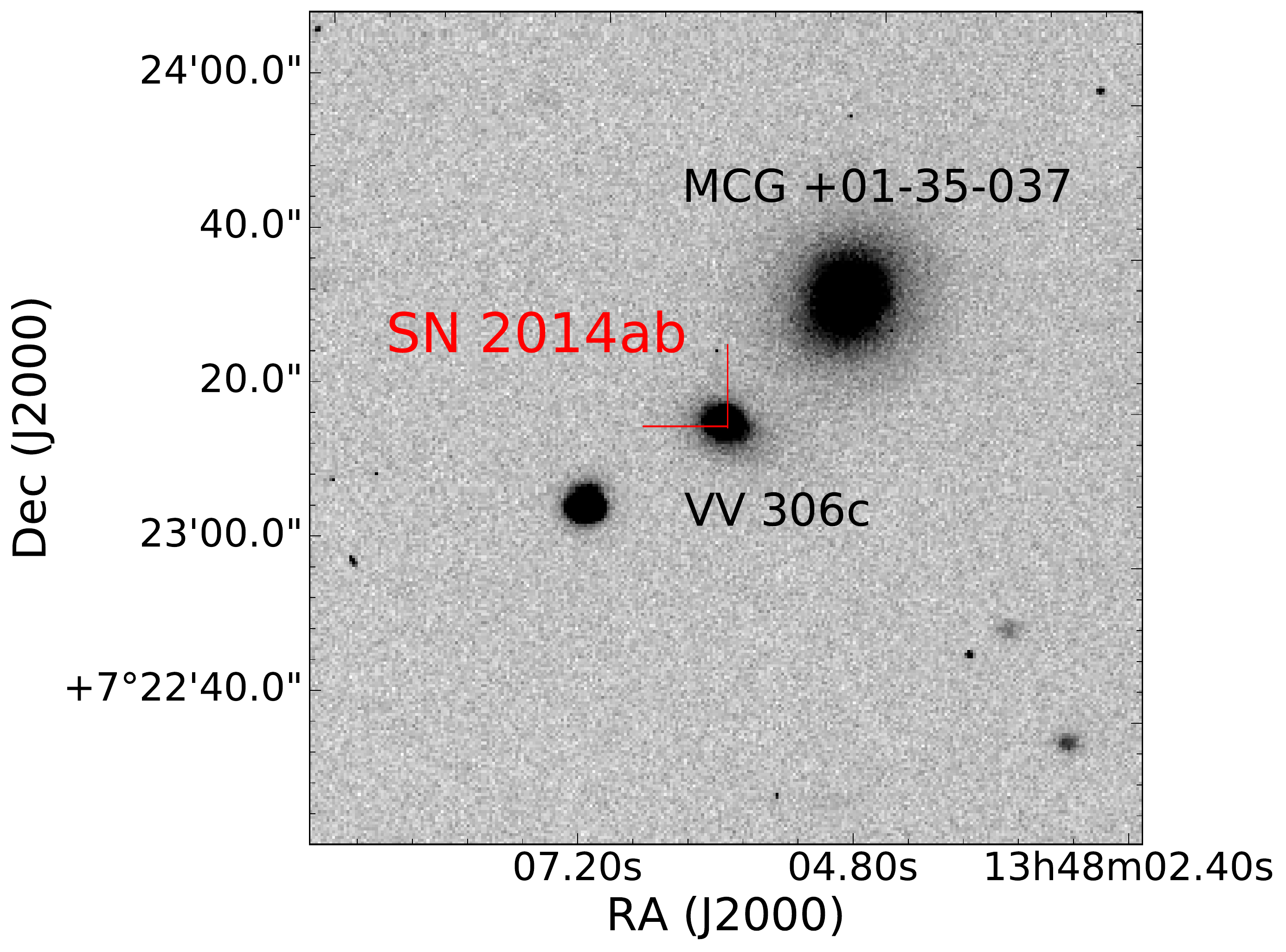}
\includegraphics[width=0.4\textwidth,clip=true,trim=0cm 0cm 0cm 0cm]{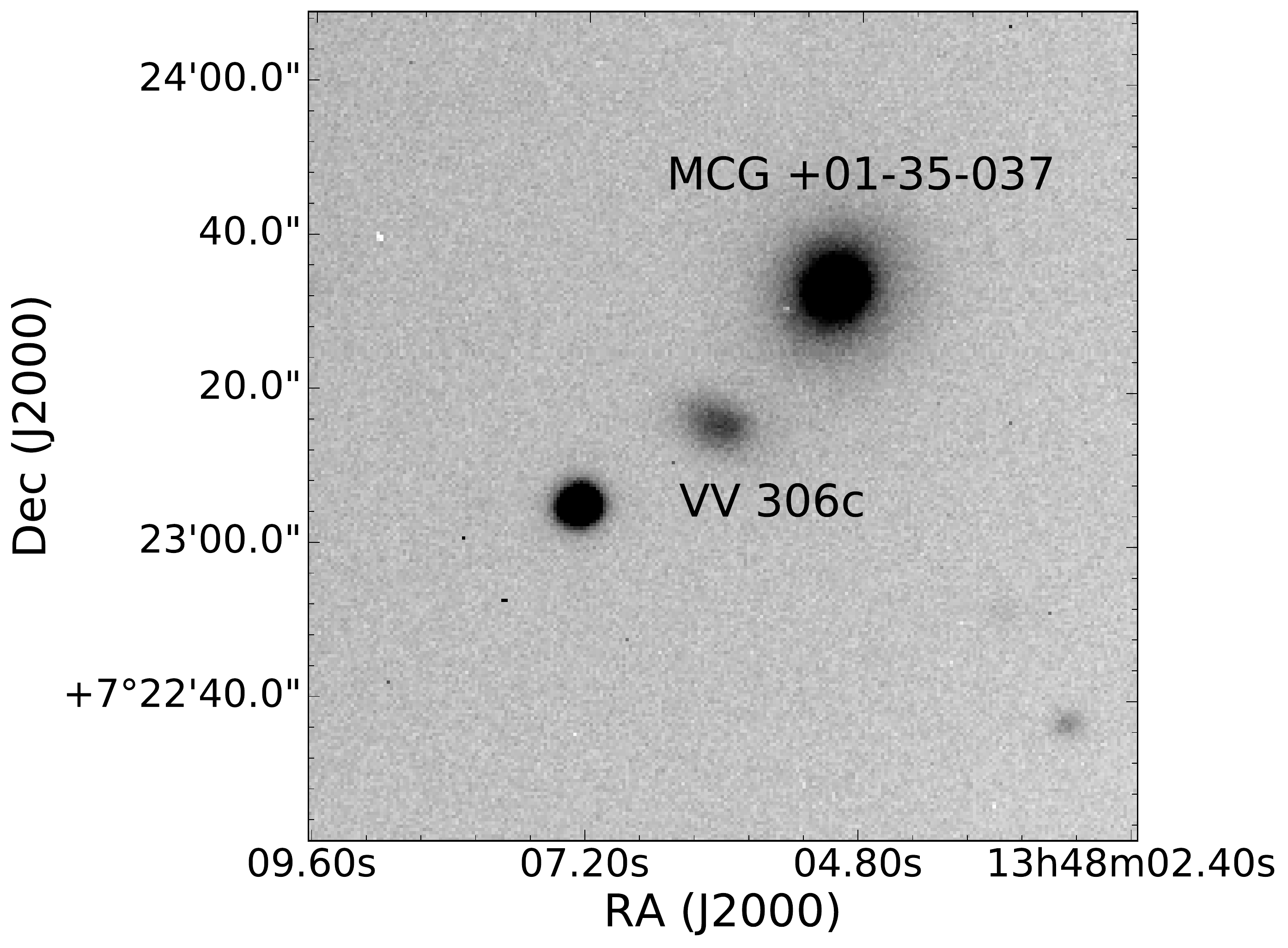}
\caption{Left: $R$-band image of VV~306c, the host galaxy of SN~2014ab, taken with the Lick Nickel 1\,m telescope with the SN present (day 72).  The red crosshairs indicate the location of SN~2014ab  Right: $R$-band image of VV~306c taken with the Kuiper telescope after the SN had faded beyond detectability (day 893).}
\label{fig:SN2014abfinder}
\end{figure*}

In this paper, we analyse the Type IIn SN~2014ab.  SN~2014ab was discovered by the Catalina Sky Survey (CSS) on 2014 Mar. 9.43 (UT dates are used in this paper) at an apparent $V$-band magnitude of 16.4 ($M_{V} = -19.0$\,mag) \citep{2009ApJ...696..870D,2014CBET.3826....1H}, located in the galaxy VV~306c (redshift $z = 0.023203$; \citealt{1959VV....C......0V}).\footnote{We note that the host galaxy is named VV~306c according to the SIMBAD database.  \citealt{2020arXiv200610198M} stated that the host galaxy of SN~2014ab is MCG +01-35-037, but this galaxy is the larger one located $\sim 20''$  to the northwest (see Fig. 1).}  The SN is located at $\alpha\mathrm{(J2000)} = 13^\mathrm{h}48^\mathrm{m}06^\mathrm{s}.05$, $\delta\mathrm{(J2000)} = +07\degree23^{\prime}16^{\prime\prime}.12$.  We adopt a Milky Way extinction along the line of sight of $A_V=0.083\,\mathrm{mag}$ ($E_{B-V} = 0.027 \,\mathrm{mag};\,$\citealt{2011ApJ...737..103S}), and a redshift-based distance [which assumes H$_0=73\,\mathrm{km\,s^{-1}\,Mpc^{-1}}$ \citep{2005ApJ...627..579R} and takes into account influences from the Virgo cluster, the Great Attractor, and the Shapley supercluster] of $105.7\pm 7.4\,\mathrm{Mpc}$ from the NASA/IPAC Extragalactic Database\footnote{The NASA/IPAC Extragalactic Database (NED) is operated by the Jet Propulsion Laboratory, California Institute of Technology, under contract with the National Aeronautics and Space Administration (NASA; \url{http://ned.ipac.caltech.edu}).}.  $R$-band images with and without the SN are shown in Figure \ref{fig:SN2014abfinder}.  An infrared (IR) spectrum acquired with the 6.5\,m Magellan Baade Telescope at Las Campanas Observatory with the FoldedPort Infrared Echellette (FIRE; 800--2,500\,nm) on 2014 Mar. 10.25 revealed features indicative of a SN~IIn more than a month past maximum brightness \citep{2014CBET.3826....1H}.  Additionally, a visual-wavelength spectrum (3985--9315\,\AA, 18\,\AA\ resolution) was obtained by the Public ESO Spectroscopic Survey for Transient Objects (PESSTO) with the ESO New Technology Telescope at La Silla on 2014 Mar. 9 using the EFOSC2 and Grism 13 \citep{2014ATel.5968....1F}.  PESSTO reports a Type~IIn classification with narrow Balmer emission lines, shallow and broad P-Cygni absorption, broad Ca near-IR triplet emission, and broad P-Cygni Na~\textsc{i}~D \citep{2014ATel.5968....1F}.  Here, we present results for SN~2014ab based on five epochs of spectropolarimetry, 31 epochs of spectroscopy, and photometry spanning 4039 days.

As our paper was in the final stage of preparation, an independent analysis of photometry and spectra of the same SN appeared in a preprint \citep{2020arXiv200610198M}.  We briefly comment on similarities and differences between our analysis and that of \citealt{2020arXiv200610198M} in the final section of this paper.

\section{Observations}

\subsection{Explosion Date and Pre-SN Activity}
Unfortunately, SN~2014ab was discovered after peak brightness and the explosion date is poorly constrained.  Prediscovery CSS imaging of the host galaxy of SN~2014ab places rough limits on the pre-explosion activity and the explosion date.  The object was detected in CSS images 56 days prior to the discovery announcement image.  We use this CSS detection date (2014 Jan. 12.4755, MJD = 56669.4755) as the true discovery date of SN~2014ab and adopt this as day 0 throughout our analysis.  CSS images of the region were taken spanning 3203 days before discovery up until 836 days after discovery (Fig. \ref{fig:fullCSS}), with no indication of any other transient events (i.e., no precursor outbursts; see \citealt{2015MNRAS.450..246B}) occurring at the location of the SN.  The brightest observed magnitude ($m_V = 15.84$, $M_V = -19.54$\,mag) occurred in the CSS image taken 56 days prior to the discovery announcement image.  CSS images taken 212 days prior to the discovery announcement image show no detection of the transient, and we have no photometry at the location of the SN between days $-212$ and 0.  It is during this time period that the explosion occurred, most likely at least tens of days before discovery.  

\label{sec:Obs}
\subsection{Photometry}
\begin{figure}
\centering
\includegraphics[width=0.5\textwidth,clip=true,trim=0cm 0cm 0cm 0cm]{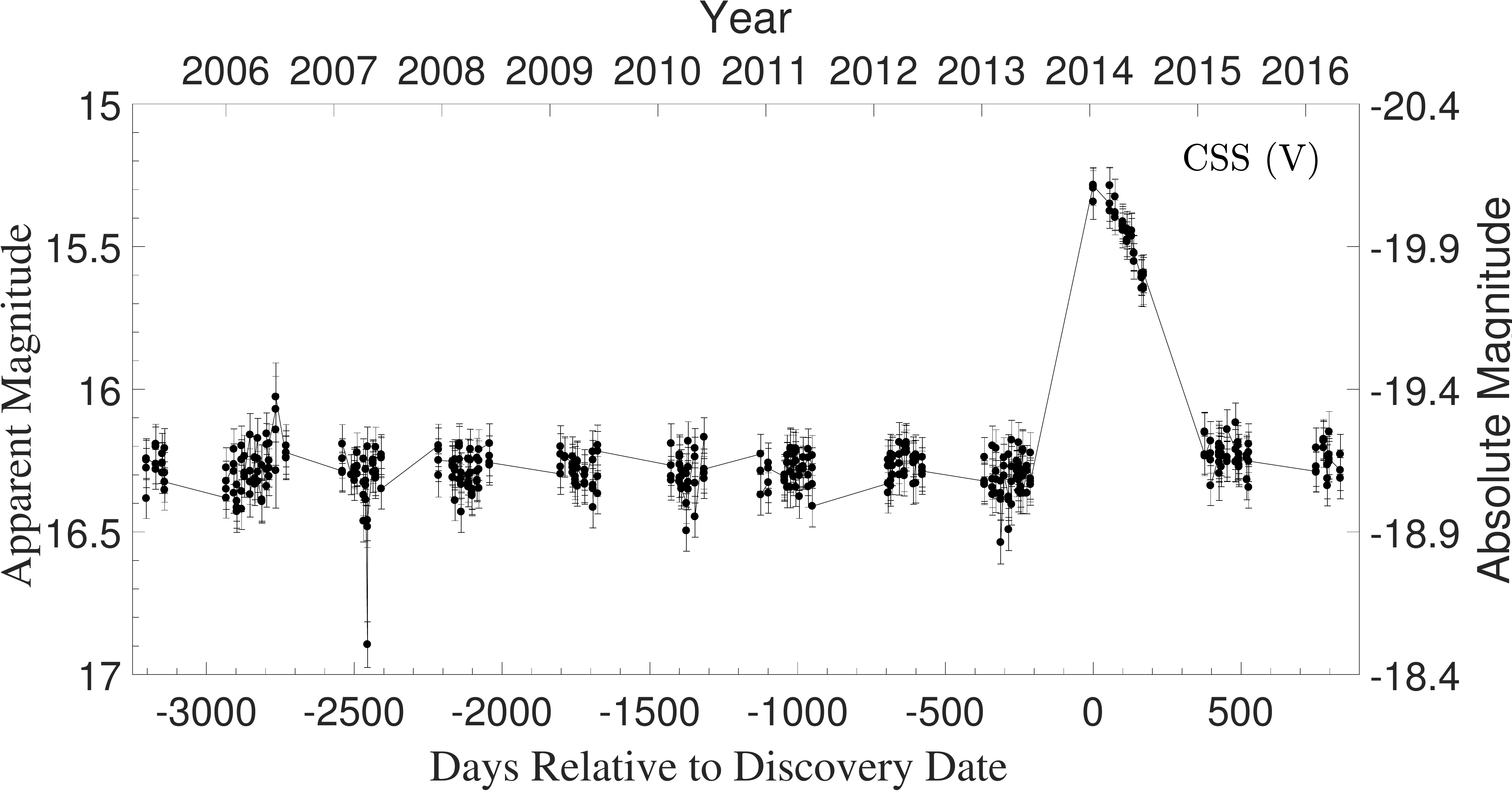}
\caption{The full range of CSS unfiltered (similar to $V$-band) photometric observations of SN~2014ab with light from its host galaxy, VV~306c, included in the aperture.  No pre-SN outburst is evident in these data.  The black line connecting the photometric observations is included to guide the eye and does not reflect any sort of model.}
\label{fig:fullCSS}
\end{figure}

We retrieved the $V$-band light curve from the CSS which spans a total of 4039 days, shown in Figure \ref{fig:fullCSS}.  As this light curve is not template-subtracted, it includes significant amounts of host-galaxy light.  We estimate the median flux in the host galaxy in the CSS images when no SN is present and subtract this host flux from the rest of the CSS light curve to estimate the SN flux.  The shape of the resulting light curve is very similar to our 1\,m Nickel telescope (at Lick Observatory) $V$-band light curve, though the fact that the CSS light curve suggests a systematically brighter SN is likely due to residual contamination from the host galaxy and the size of the region upon which photometry was performed.  We also obtained a late-time (day 893) $V$-band image of the host galaxy (shown in Fig. \ref{fig:SN2014abfinder}) with the Mont4K CCD Imager on the Kuiper telescope.

\begin{figure*}
\centering
\includegraphics[width=1\textwidth,clip=true,trim=0cm 0cm 0cm 0cm]{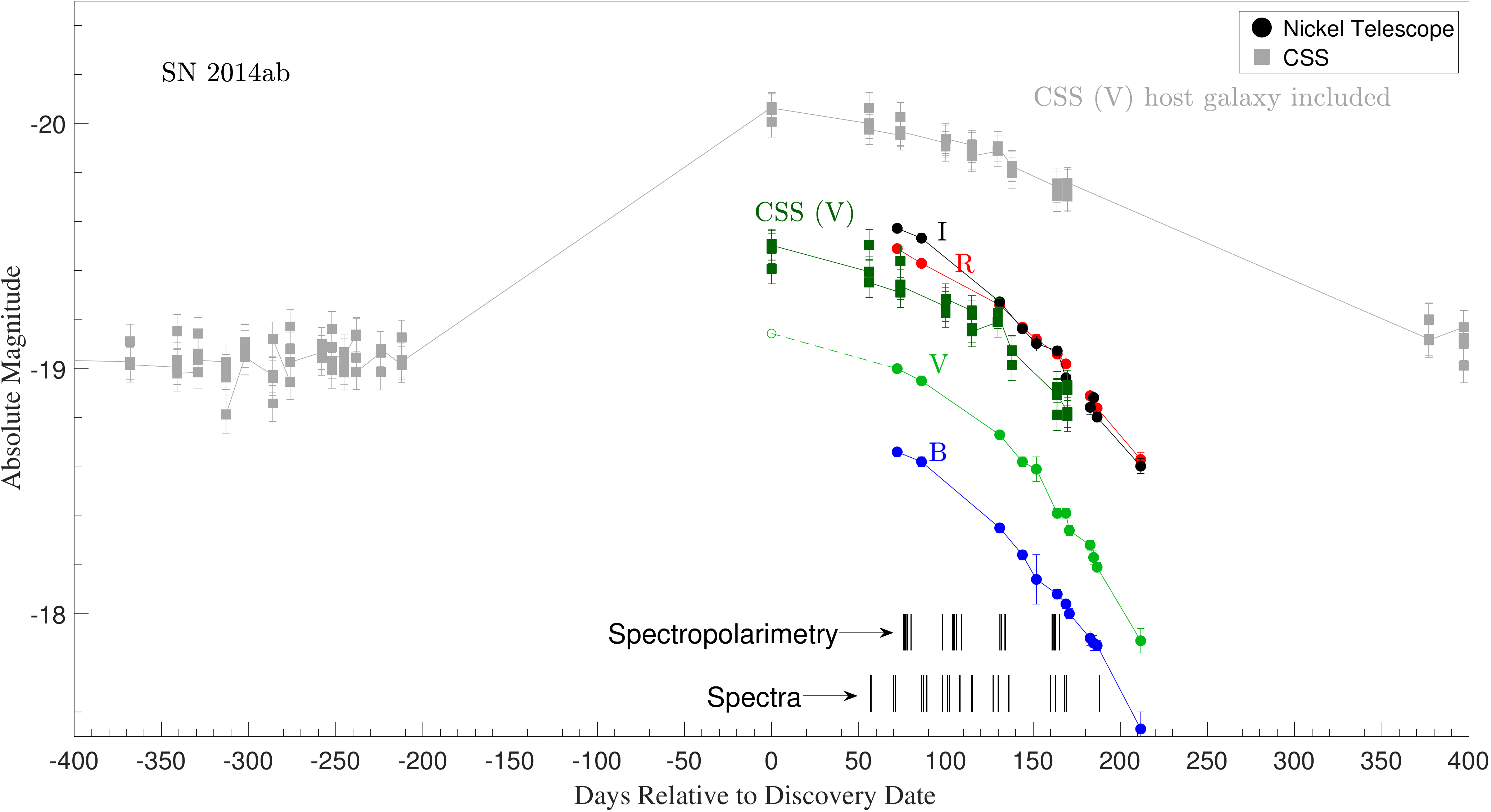}
\caption{Photometric observations of SN~2014ab taken by the Lick Nickel 1\,m telescope ($BVRI$) and the CSS (unfiltered, but similar to $V$-band).  CSS data that include host-galaxy light (from Fig. \ref{fig:fullCSS}) are shown in grey, while host-flux-subtracted CSS unfiltered measurements are shown in dark green.  Nickel $V$-band data does not extend back to the discovery date, so we have shown a projected peak magnitude value (based on extrapolating the shape of the CSS unfiltered light curve) with an unfilled green circle connected by a dotted line.  Dates on which we also obtained spectra and spectropolarimetry are marked by black dashes. Dates are relative to the adopted discovery date of 2014 Jan. 12.4755.}
\label{fig:photometry}
\end{figure*}

$B$, $V$, $R$, and $I$ images of SN~2014ab obtained with the Nickel telescope were reduced using a custom pipeline \citep{2010ApJS..190..418G}.  Image-subtraction procedures performed by the ISIS package \citep{1998ApJ...503..325A} were applied in order to remove the host-galaxy light, using template images obtained on 2016 Jan. 26, after the SN had faded below our detection limit.  Point-spread-function (PSF) photometry was then obtained using DAOPHOT \citep{1987PASP...99..191S} from the IDL Astronomy User’s Library\footnote{http://idlastro.gsfc.nasa.gov/}.  Several nearby stars were chosen from the SDSS catalog for calibrating Nickel data.  Their magnitudes were first transformed into the Landolt system using the empirical prescription presented by Robert Lupton\footnote{http://www.sdss.org/dr7/algorithms/sdssUBVRITransform.html \\ \#Lupton2005}, then transformed to the Nickel natural system.  Apparent magnitudes were all measured in the Nickel natural system, and the final results were transformed to the standard system using local calibrators and colour terms of ``Nickel2" (see Table 1 from \citealt{2019MNRAS.490.3882S}.  We display the Nickel telescope light curve in Figure \ref{fig:photometry} alongside an additional unfiltered (similar to $V$-band) light curve from the CSS.

\subsection{Spectroscopy}
We obtained optical spectra with a variety of telescopes instruments over the course of $\sim 200$ days after the detection of SN~2014ab. Our 36 spectra were taken on 34 different nights with the Kast double spectrograph \citep{1993MillerStone} on the Shane 3\,m telescope at Lick Observatory; EFOSC2 on the 3.58\,m ESO-NTT \citep{1984Msngr..38....9B}; X-shooter on the 8.2\,m VLT \citep{2011A&A...536A.105V}; the FLOYDS spectrograph on the 2\,m Las Cumbres Observatory telescopes as part of the LCO Supernova Key Project \citep{2013PASP..125.1031B}; the Bluechannel (BC) Spectrograph on the 6.5\,m Multiple Mirror Telescope (MMT); and the Spectropolarimeter (SPOL) on the 1.54\,m Kuiper telescope, the 2.3\,m Bok telescope, and the the 6.5\,m MMT.  All spectra were taken with the long slit at the parallactic angle \citep{1982PASP...94..715F}.  The spectroscopic observations are detailed in Table \ref{tab:spectra}. 

\begin{figure*}
\centering
\includegraphics[width=1\textwidth,clip=true,trim=0cm 0cm 0cm 0cm]{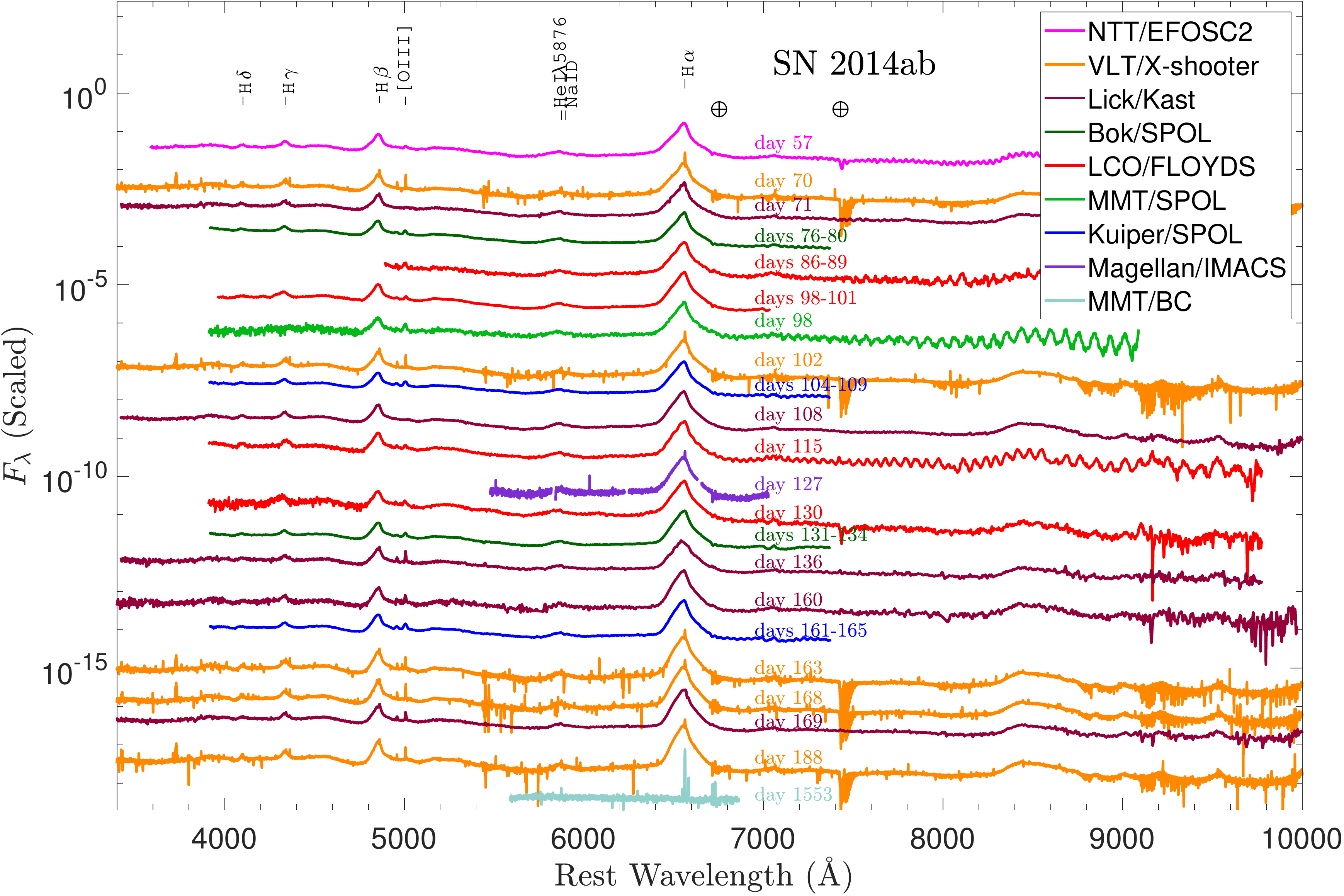}
\caption{Visual-wavelength spectra of SN~2014ab.  All spectra were corrected for reddening using $E_{B-V} = 0.083$\,mag, which accounts for the total Milky Way and estimated host-galaxy reddening (see text). All spectra have also been corrected for host-galaxy redshift, and scaled for clarity (see Table \ref{tab:spectra}).  Additional small wavelength corrections were applied so that the narrow component of H$\alpha$ consistently matches the rest-frame wavelength of H$\alpha$. Telluric features were not removed from some of the spectra and are marked with an Earth symbol; those longward of $\sim 9000$\,\AA\ are not marked.  Days since discovery date indicated.}
\label{fig:allspectra}
\end{figure*}

\begin{figure*}
\centering
\includegraphics[width=1\textwidth,clip=true,trim=0cm 0cm 0cm 0cm]{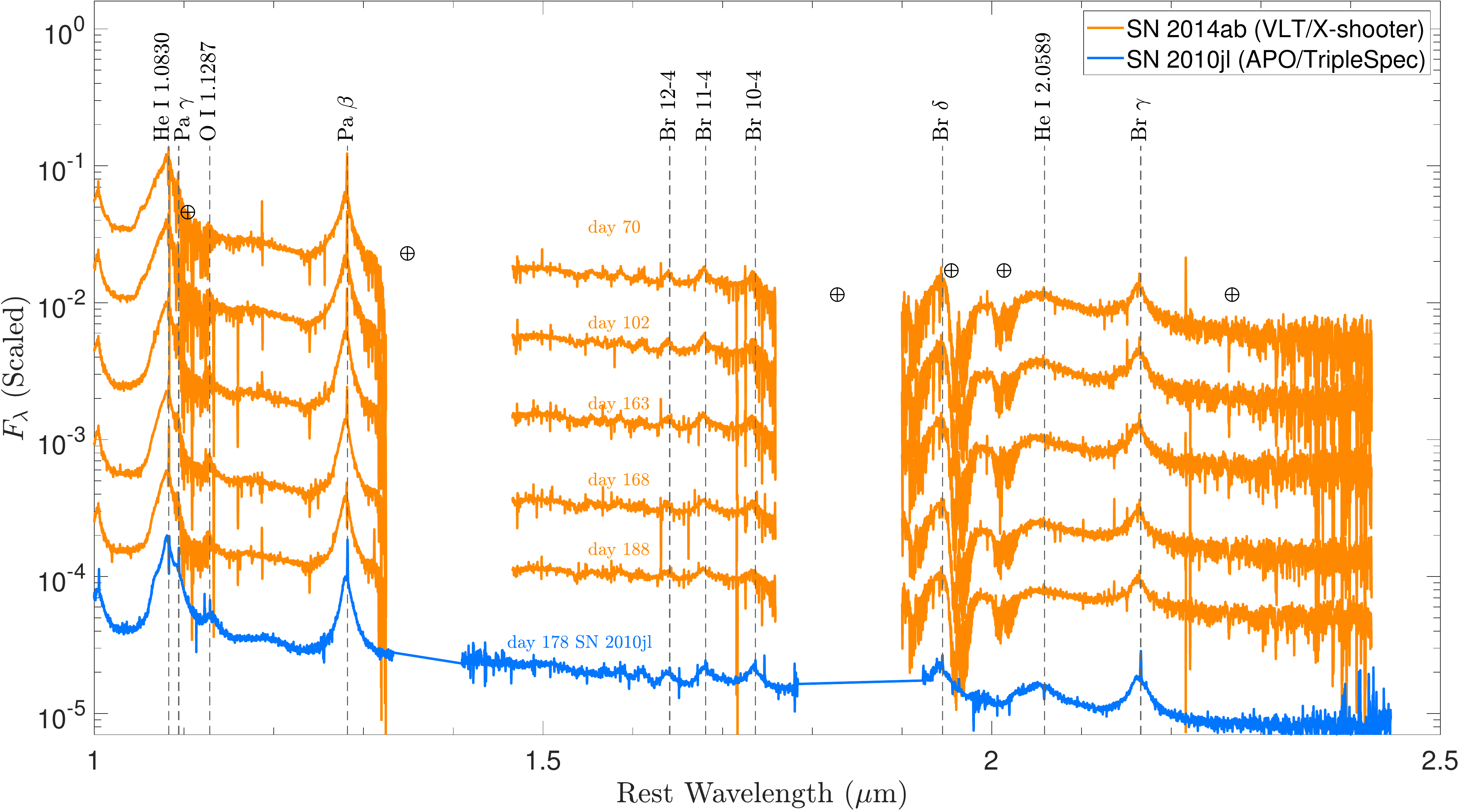}
\caption{Near-IR spectra of SN~2014ab.  All spectra were corrected for reddening using $E_{B-V} = 0.083$\,mag, which accounts for the total Milky Way and estimated host-galaxy reddening (see text). All spectra have also been corrected for host-galaxy redshift, and scaled for clarity (see Table \ref{tab:spectra}).  Additional small wavelength corrections were applied so that the narrow component of H$\alpha$ consistently matches the rest-frame wavelength of H$\alpha$, though H$\alpha$ is not shown in this figure. Telluric features were not removed and are marked with an Earth symbol. We also plot a spectrum of SN~2010jl 178 days after earliest detection for comparison \citep{2015ApJ...801....7B}.}
\label{fig:IRspectra}
\end{figure*}

Standard spectral reduction procedures were followed for all of the spectra (except SPOL polarization data; see \S \ref{sec:Obs_SPOL}).
The reduced optical spectra are shown in Figure \ref{fig:allspectra}. While we focus our analysis on the optical spectra, the X-shooter spectra also contain near-IR data, plotted in Figure \ref{fig:IRspectra}.

\begin{table*}
\begin{minipage}{200mm}
\caption{Spectroscopic Observations of SN 2014ab}
\label{tab:spectra}
\begin{tabular}{ccccccc}
  \hline
Year-Month-Day	&	Day\tablenotemark{a}	&	Telescope/Instrument & Wavelength Range ({\AA}) & $\sim R\,(\lambda/\Delta\lambda)$\\
	\hline
2014-03-10	&	57 &	ESO-NTT/EFOSC2	& 3585-9065 & 350\\
2014-03-23 	&   70 & ESO-VLT/X-shooter 	& 2923-24241 & 3500\\
2014-03-24	&	71 &	Lick/Kast		& 3421-9775 & 600\\
2014-03-29	&	76 &	Bok/SPOL		& 3911-7373 & 200\\
2014-03-30	&	77 &	Bok/SPOL		& 3909-7375 & 200\\
2014-03-31	&	78 &	Bok/SPOL		& 3912-7375 & 200\\
2014-04-02	&	80 &	Bok/SPOL 		& 3910-7377 & 200\\
2014-04-08	&	86 & LCO/FLOYDS			& 4888-11003 & 550\\
2014-04-09	&	87 & LCO/FLOYDS			& 4892-9096 & 550\\
2014-04-11	&	89 & LCO/FLOYDS			& 4886-9770 & 550\\
2014-04-20	&	98 & LCO/FLOYDS			& 3960-7036 & 550\\
2014-04-20	&	98 & MMT/SPOL			& 3910-9090 & 400\\
2014-04-23	&	101 & LCO/FLOYDS		& 3913-9779 & 550\\
2014-04-24 	&   102 & ESO-VLT/X-shooter & 2923-24241 & 3500\\		
2014-04-26	&	104 &	Kuiper/SPOL		& 3912-7375 & 200\\
2014-04-27	&	105 &	Kuiper/SPOL		& 3911-7372 & 200\\
2014-04-28	&	106 &	Kuiper/SPOL		& 3908-7375 & 200\\
2014-04-30	&	108 &	Lick/Kast		& 3420-10261 & 600\\
2014-05-01	&	109 &	Kuiper/SPOL		& 3910-7377 & 200\\
2014-05-07	&	115 & LCO/FLOYDS		& 3910-9774 & 550\\	
2014-05-19  &   127 & Magellan/IMACS	& 5477-7028 & 1600\\
2014-05-22	&	130 & LCO/FLOYDS		& 3911-9776 & 550\\	
2014-05-23	&	131 &	Bok/SPOL		& 3911-7379 & 200\\
2014-05-24	&	132 &	Bok/SPOL		& 3912-7375 & 200\\
2014-05-26	&	134 &	Bok/SPOL		& 3914-7379 & 200\\
2014-05-28	&	136 &	Lick/Kast		& 3421-9774 & 600\\
2014-06-21	&	160 &	Lick/Kast		& 3382-9970 & 600\\
2014-06-22	&	161 &	Kuiper/SPOL		& 3911-7379 & 200\\
2014-06-23	&	162 &	Kuiper/SPOL		& 3910-7376 & 200\\
2014-06-24	&	163 &	Kuiper/SPOL		& 3911-7375 & 200\\
2014-06-24 	&   163 & ESO-VLT/X-shooter	& 2923-24239 & 3500\\
2014-06-26	&	165 &	Kuiper/SPOL		& 3909-7375 & 200\\
2014-06-29 	&   168 & ESO-VLT/X-shooter	& 2923-24239 & 3500\\
2014-06-30	&	169 &	Lick/Kast		& 3373-10266 & 600\\
2014-07-19	&   188 & ESO-VLT/X-shooter	& 2923-24239 & 3,500\\
2018-04-14  &   1553& MMT/BlueChannel   & 5588-6863 & 3310\\
	\hline
	\end{tabular}
	\tablenotetext{a}{Days since adopted discovery date (2014 Jan. 12.4755 UT).}
\end{minipage}
\end{table*}

\subsection{Spectropolarimetry}
\label{sec:Obs_SPOL}
Spectropolarimetric observations of SN~2014ab were obtained using the CCD Imaging/Spectropolarimeter \citep[SPOL; ][]{1992ApJ...398L..57S} on the 2.3\,m Bok, 1.54\,m Kuiper, and 6.5\,m MMT telescopes.  A $5''$ slit was used at the Bok and Kuiper telescopes, while a $3''$ slit was used at the MMT.  Observation and data reduction procedures were followed as in \citealt{2018MNRAS.475.1104B}, except using a wavelength range of 4000--7550\,\AA.  Nine $q$ and $u$ sequences were acquired at the Bok telescope, 11 at the Kuiper telescope, and 2 at the MMT.  Each set of sequences was then combined by epoch for a higher signal-to-noise ratio.  

Hiltner 960 and VI~Cyg~12 were used as polarimetric standards \citep{1992AJ....104.1563S} to obtain the instrumental polarization angle for SPOL at the Bok and MMT telescopes.  The discrepancy between the measured and the expected position angle was $<0.2\degree$ for each of the polarimetric standard stars.  BD+28$\degree$4211 was used as an unpolarized flux standard to ensure that the instrumental polarization for SPOL was $<0.1$\% for each epoch \citep{1990AJ.....99.1621O}.  We also use unpublished spectropolarimetry of SN~2010jl \citep{inprep} obtained by the SN Spectropolarimetry Project to compare to our spectropolarimetry of SN~2014ab.

\section{Results}
\label{sec:Res}

\subsection{Extinction and Reddening}
\label{sec:Ext}
\begin{figure}
\centering
\includegraphics[width=0.48\textwidth,clip=true,trim=0cm 0cm 0cm 0cm]{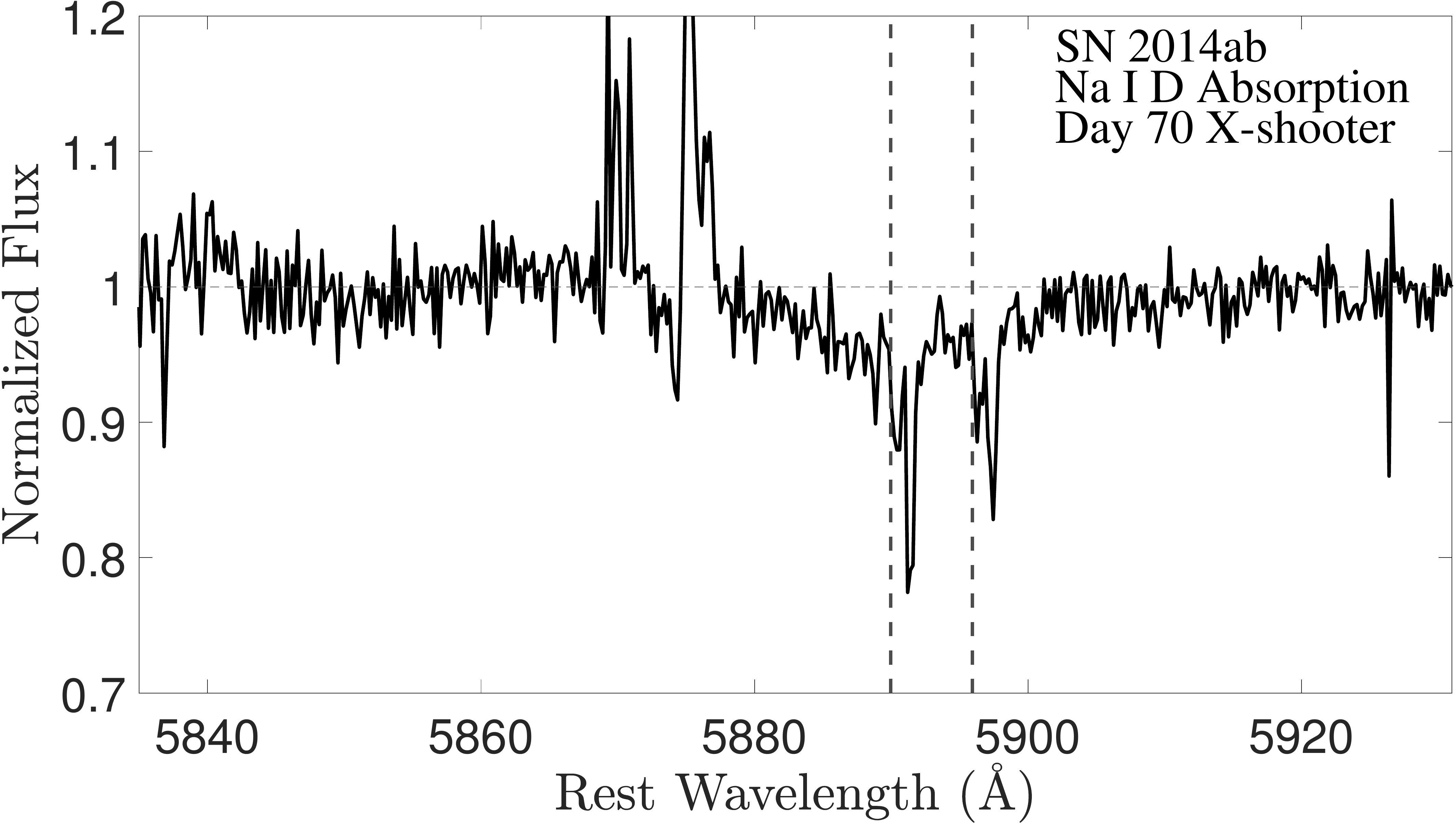}
\caption{An $R=3500$ X-shooter spectrum taken on day 70 showing the sodium doublet absorption which we use to constrain the host-galaxy extinction and interstellar polarization.  The spectrum has been redshift-corrected using the average redshift of the host galaxy and then aligned to the narrow component of H$\alpha$ emission.  Vertical dashed lines show the rest-frame wavelengths of the Na~\textsc{i}~D lines.  The observed Na~\textsc{i}~D absorption lines are offset slightly ($\sim 54$\,km\,$\mathrm{s^{-1}}$) from their rest wavelengths because of host-galaxy rotation.} 
\label{fig:NaID}
\end{figure}

We use the strength of Na~\textsc{i}~D absorption to evaluate the local reddening along the line of sight to SN~2014ab within its host galaxy.  The strength of the narrow absorption lines ofNa~\textsc{i}~D $\lambda\lambda$5890 (D2), 5896 (D1) correlates with the interstellar dust extinction present along a particular line of sight.  While this relation does not perform well with low-resolution spectra \citep{2011MNRAS.415L..81P}, it can be used with moderate-resolution spectra when the Na~\textsc{i}~D2 line is not saturated and the doublet is not blended \citep{2012MNRAS.426.1465P}.  \citet{2013ApJ...779...38P} found that the sodium doublet absorption for one-fourth of their sample of SNe~Ia was stronger than expected for dust-extinction values estimated from SN colour.  In our moderately high-resolution ($R = 3500$) spectrum on day 70, we measure the equivalent widths for the D1 and D2 lines ($\lambda 5896$ and $\lambda 5890$, respectively, in the host-galaxy rest frame) to be $0.21 \pm 0.03$ \r{A} and $0.31 \pm 0.03$ \r{A}, respectively (see Fig. \ref{fig:NaID}).  Based on these equivalent widths, the relation provided by \citet{2012MNRAS.426.1465P} suggests that we have additional extinction along the line of sight caused by the host galaxy of $A_V \approx 0.18$\,mag (assuming $A_V = 3.08 E_{B-V}$; \citealt{1992ApJ...395..130P}) or $A_R \approx 0.14$\,mag.\footnote{Other attempts have been made to connect the equivalent width of the absorption in the sodium doublet to extinction \citep{1994AJ....107.1022R,1997A&A...318..269M,2003fthp.conf..200T,2012MNRAS.426.1465P}.}.  If we choose the model (described in Footnote 5) with the highest estimated extinction we would have $A_V = 0.55$\,mag instead.  We adopt a total Milky Way \citep{2011ApJ...737..103S} plus host-galaxy extinction of $A_V = 0.26$\,mag ($E_{B-V}=0.083$\,mag) or $A_R = 0.21$\,mag.  Figure \ref{fig:allspectra} shows spectra dereddened by $E_{B-V}=0.083$\,mag.

\subsection{Light Curve}
\label{sec:Res:LC}
Figure \ref{fig:photometry} \, shows the light curve of SN~2014ab obtained by the Nickel telescope and CSS.  The absolute magnitudes shown have been adjusted for Milky Way and host-galaxy reddening (determined from Na~\textsc{i}~D line strengths as discussed in \S \ref{sec:Ext}) and for the distance modulus of $\mu= 35.1$\,mag based on a distance to the host galaxy of $105.7\pm 7.4\,\mathrm{Mpc}$.

CSS photometry of the host galaxy prior to the initial discovery announcement suggests that SN~2014ab initially brightened sometime between days $-212$ and 0 relative to the discovery date.  We measure SN~2014ab to have the brightest observed magnitude at $M_V =-19.54$\,mag on the first CSS detection date, so the peak luminosity may have occurred some time before then.  Even our earliest spectra resemble those of SN~2010jl \citep{2012AJ....143...17S} more than 50 days after peak, suggesting that SN~2014ab was already well past maximum brightness when discovered.  For this reason, all dates are cited relative to the CSS first detection date, which may be several weeks or months after the SN exploded.

Because the CSS photometry included host-galaxy light and is unfiltered rather than true $V$-band data, we favour using the Nickel data for a constraint on the peak $V$-band magnitude.  Specifically, we have estimated the magnitude the $V$-band Nickel data would have if it followed the same changes in brightness that the CSS data experienced between days 0 and 72.  This results in a projected peak $V$-band magnitude of -19.14 in the Nickel data, which we use as the $V$-band peak estimate.  Note, however, that SN~2014ab's peak magnitude occurs on the discovery date, with a gap of data in the 212 days prior, suggesting that we only have estimated a lower limit to the peak brightness of SN~2014ab.

The decline rate for SN~2014ab was $\sim 0.0034$\,mag $\mathrm{day^{-1}}$ in the CSS data and $\sim 0.0079$\,mag $\mathrm{day^{-1}}$ in the Nickel data, with the somewhat steeper decline in Nickel data reflecting the later dates that were sampled.  We compare the light-curve decline rates to those of other SNe in \S \ref{sec:Dis:LC}.

\subsection{SN Location}

We find astrometric fits to both the Nickel and Kuiper telescope images (Fig. \ref{fig:photometry}) using \url{astrometry.net} \citep{2010AJ....139.1782L}.  We then measure the location of the host galaxy from the Kuiper image and the location of the SN from the Nickel image after template subtraction using radial profile fits to a Moffat distribution \citep{1969A&A.....3..455M}.  We determine the uncertainty in the location of the centroid by replicating the noise level in each image and refitting the centroid 100 times.  The location of the SN is measured to be $\alpha\mathrm{(J2000)} = 13^\mathrm{h}48^\mathrm{m}06^\mathrm{s}.05$ and $\delta\mathrm{(J2000)} = +07\degree23^{\prime}16^{\prime\prime}.12 \pm 0.01^{\prime\prime}$, and the location of the host galaxy is measured to be $\alpha\mathrm{(J2000)} = 13^\mathrm{h}48^\mathrm{m}06^\mathrm{s}.01$ and $\delta\mathrm{(J2000)} = +07\degree23^{\prime}15^{\prime\prime}.41 \pm 0.3^{\prime\prime}$.  The difference between these is $\Delta\alpha = 0.05^\mathrm{s} \pm 0.25^\mathrm{s}$ and $\Delta\delta = 0.71^{\prime\prime} \pm 0.17^{\prime\prime}$.  $\Delta\delta$ is bigger than the uncertainty in the $\Delta\delta$, suggesting that the SN is offset from the host-galaxy nucleus by roughly $0.71^{\prime\prime}$.  Given the distance to SN~2014ab of 105.7\,Mpc, it has a projected distance from the nucleus of the host galaxy of $\sim 364$\,pc.  This indicates that SN~2014ab is not a nuclear transient like a tidal disruption event.

\subsection{Spectral Morphology}
\begin{figure}
\centering
\includegraphics[width=0.5\textwidth,clip=true,trim=0cm 0cm 0cm 0cm]{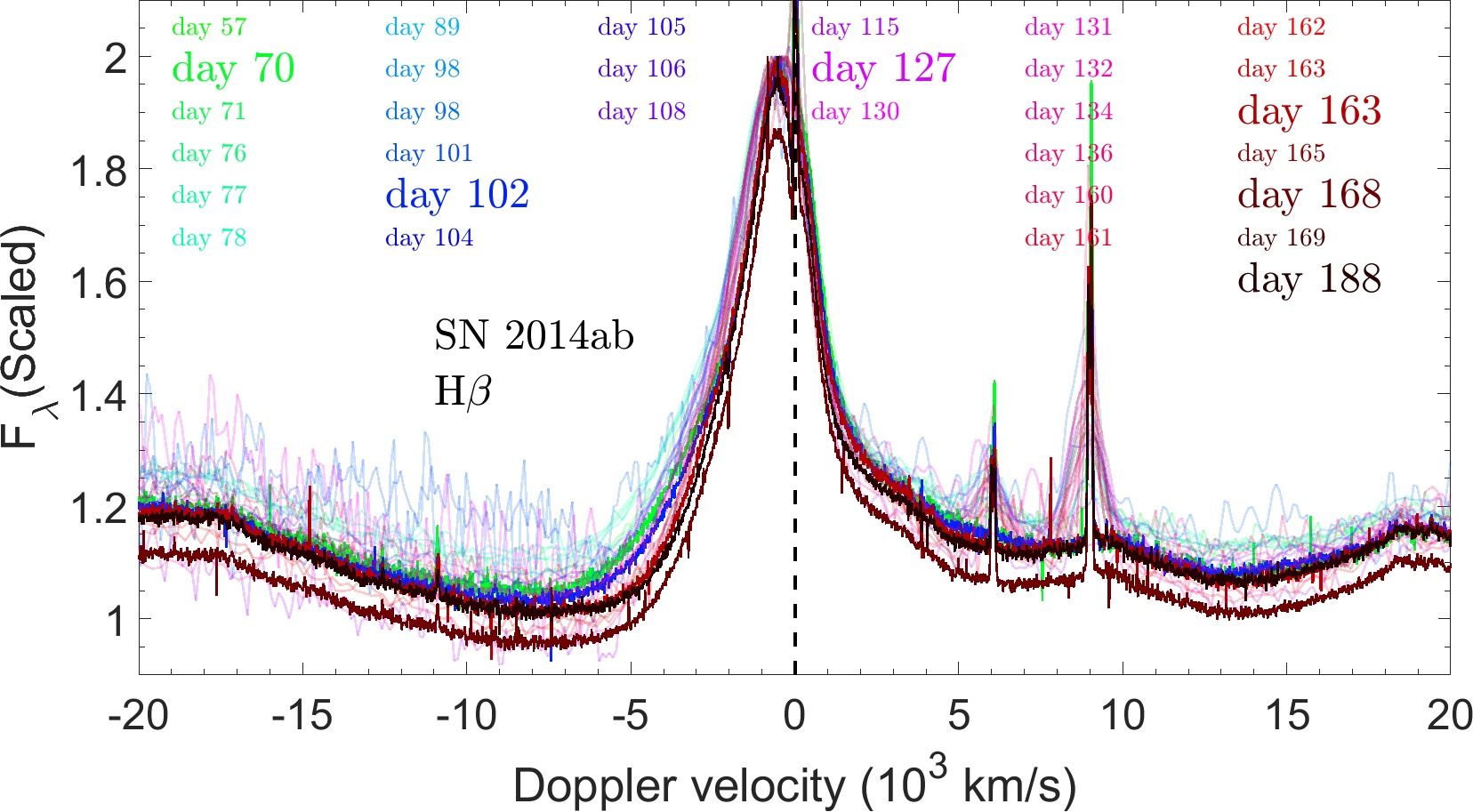}
\caption{The progression of the H$\beta$ line over the course of the $\sim 150$ days during which we obtained spectra.  Medium- and high- resolution spectra are shown fully opaque and have their dates labeled with larger text.  Because the spectra overlap heavily owing to a lack of change in the line profile, low-resolution spectra are shown transparently and have small date labels.  A broad, shallow P-Cygni absorption feature is seen, likely arising in the SN ejecta.  This absorption feature has a minimum at $v \approx -8,000$\,km\,$\mathrm{s^{-1}}$, and a blue edge out to $v \approx -18,000$\,km\,$\mathrm{s^{-1}}$.}
\label{fig:Hbeta}
\end{figure}

\begin{figure}
\centering
\includegraphics[width=0.5\textwidth,height=1\textheight,keepaspectratio,clip=true,trim=0cm 0cm 0cm 0cm]{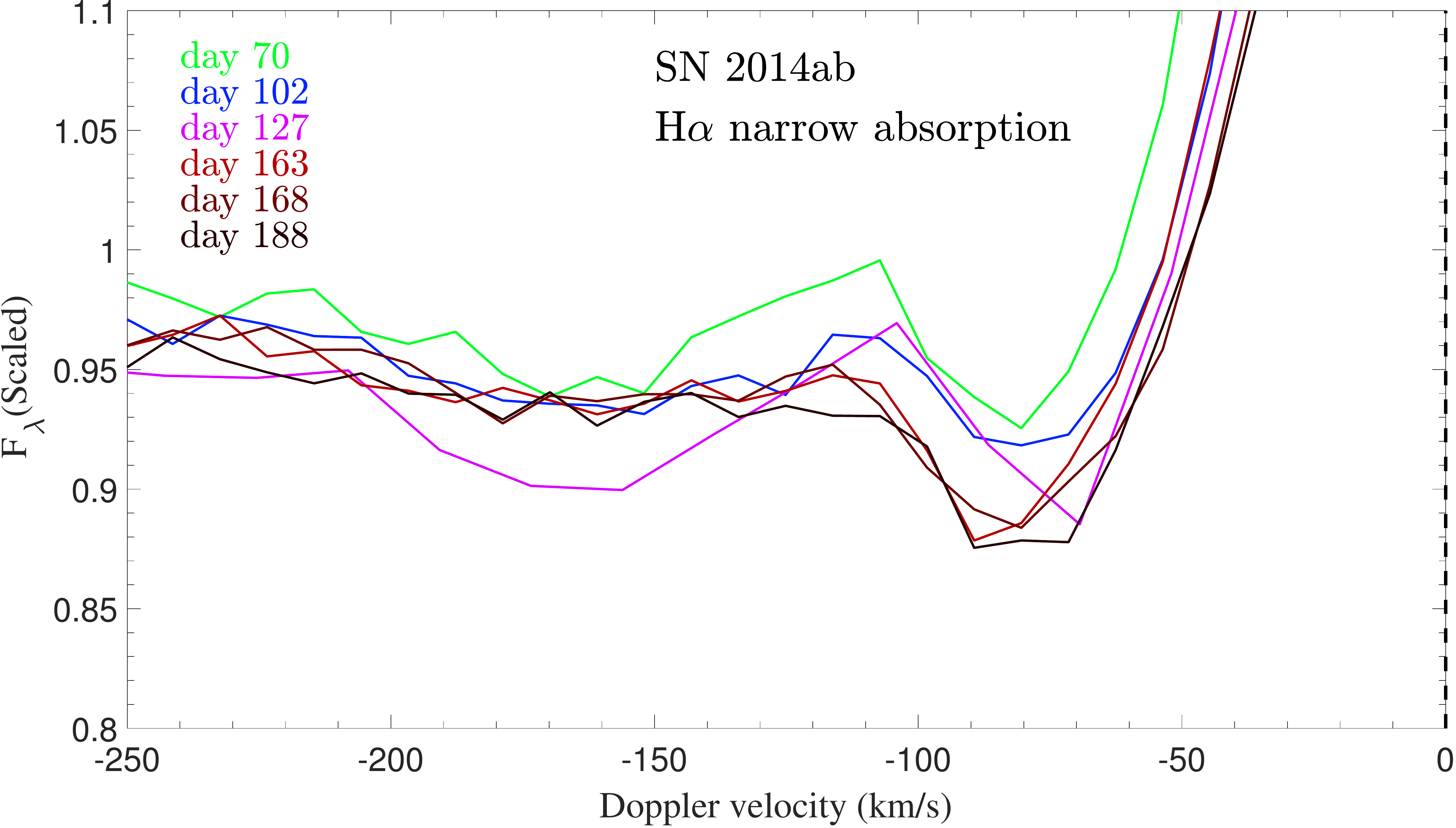}
\includegraphics[width=0.5\textwidth,height=1\textheight,keepaspectratio,clip=true,trim=0cm 0cm 0cm 0cm]{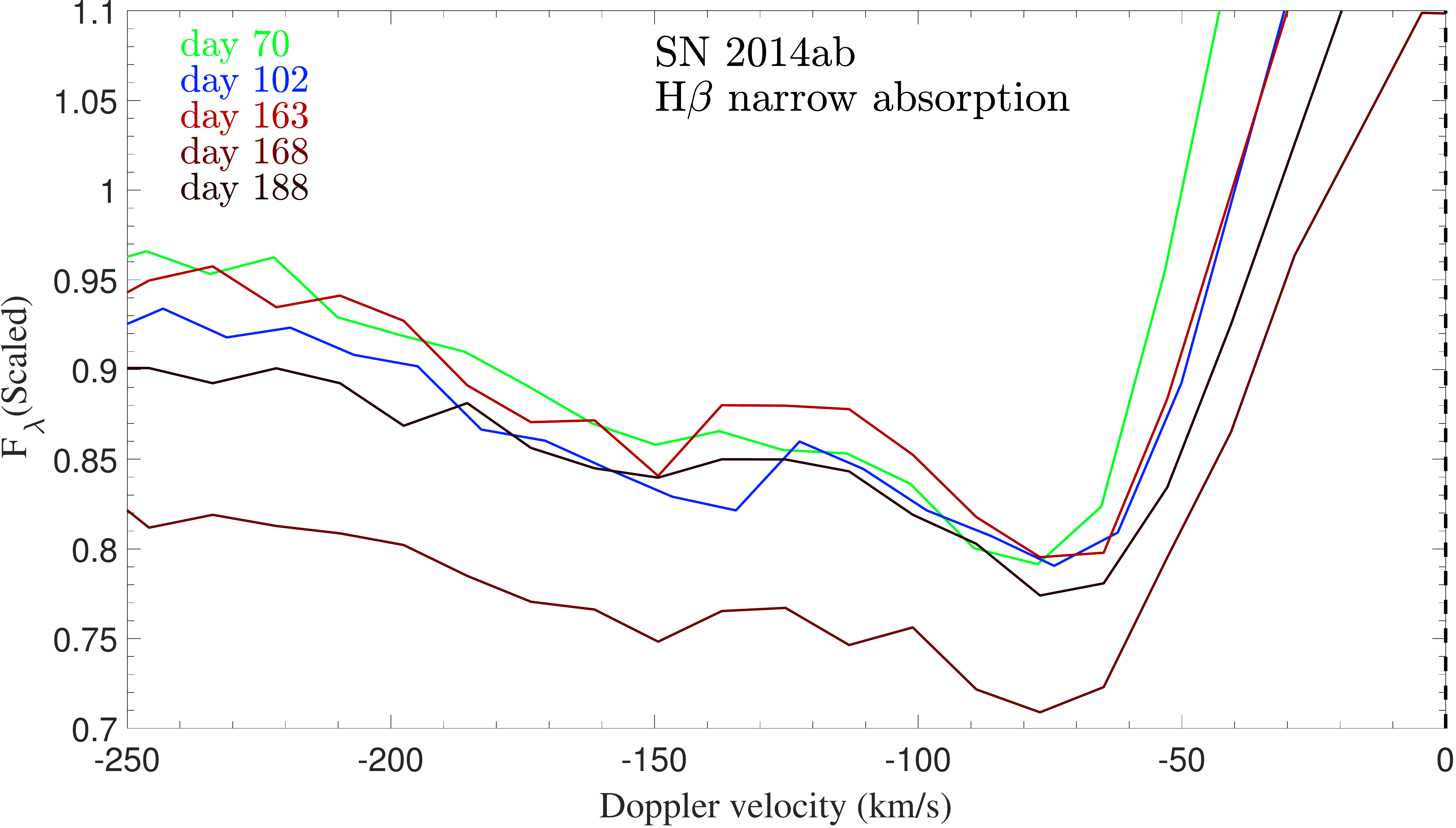}
\caption{{\it Top panel:} Narrow absorption features of H$\alpha$ ($v \approx -80$\,km\,$\mathrm{s^{-1}}$) are seen in all of our higher resolution spectra.  {\it Bottom panel:} The same for H$\beta$.}
\label{fig:HalphaHbetaNarrowAbs}
\end{figure}

We detect a variety of spectral features in the 36 different spectra.  Intermediate-width H$\alpha$ and H$\beta$ emission are the most prominent spectral features present from day 57 to day 188.  H$\beta$ shows a broad, shallow, blueshifted absorption feature with a minimum of the absorption at $v \approx -8000$\,km\,$\mathrm{s^{-1}}$, and with the blue edge of the absorption extending to about $-$17,000\,km\,s$^{-1}$ (see Fig. \ref{fig:Hbeta}).  We do not see similarly strong, broad absorption in H$\alpha$.  H$\alpha$ and H$\beta$ both exhibit narrow blueshifted absorption lines ($v \approx -80$\,km\,$\mathrm{s^{-1}}$; see Fig. \ref{fig:HalphaHbetaNarrowAbs}) in the higher resolution spectra we obtained.  The Ca~\textsc{ii} near-IR triplet (a blend of $\lambda$8498, $\lambda$8542, and $\lambda$8662) is strong in emission in SN~2014ab at all epochs that we observe.  We also detect a number of weaker emission features in many of the optical spectra: these include H$\delta$, H$\gamma$, and [O~\textsc{iii}] $\lambda\lambda$4959, 5007.  Narrow H$\alpha$ emission is distinct from the intermediate-width feature in our higher resolution spectra, but it is blended with the intermediate-width component in our lower resolution spectra.  X-shooter spectra extend into the near-IR ($\sim 2.4\mu$m,), revealing a number of emission features.  Most prominently, we detect He~\textsc{i} $\lambda$10,830, Pa$\beta$, Br$\delta$, and Br$\gamma$.  We also detect weaker emission features of presumably Pa$\gamma$, O~\textsc{i} $\lambda$11,287, Br 12-4, Br 11-4, Br 10-4, and He~\textsc{i} $\lambda$20,589.  

\begin{figure}
\centering
\includegraphics[width=0.5\textwidth,clip=true,trim=0cm 0cm 0cm 0cm]{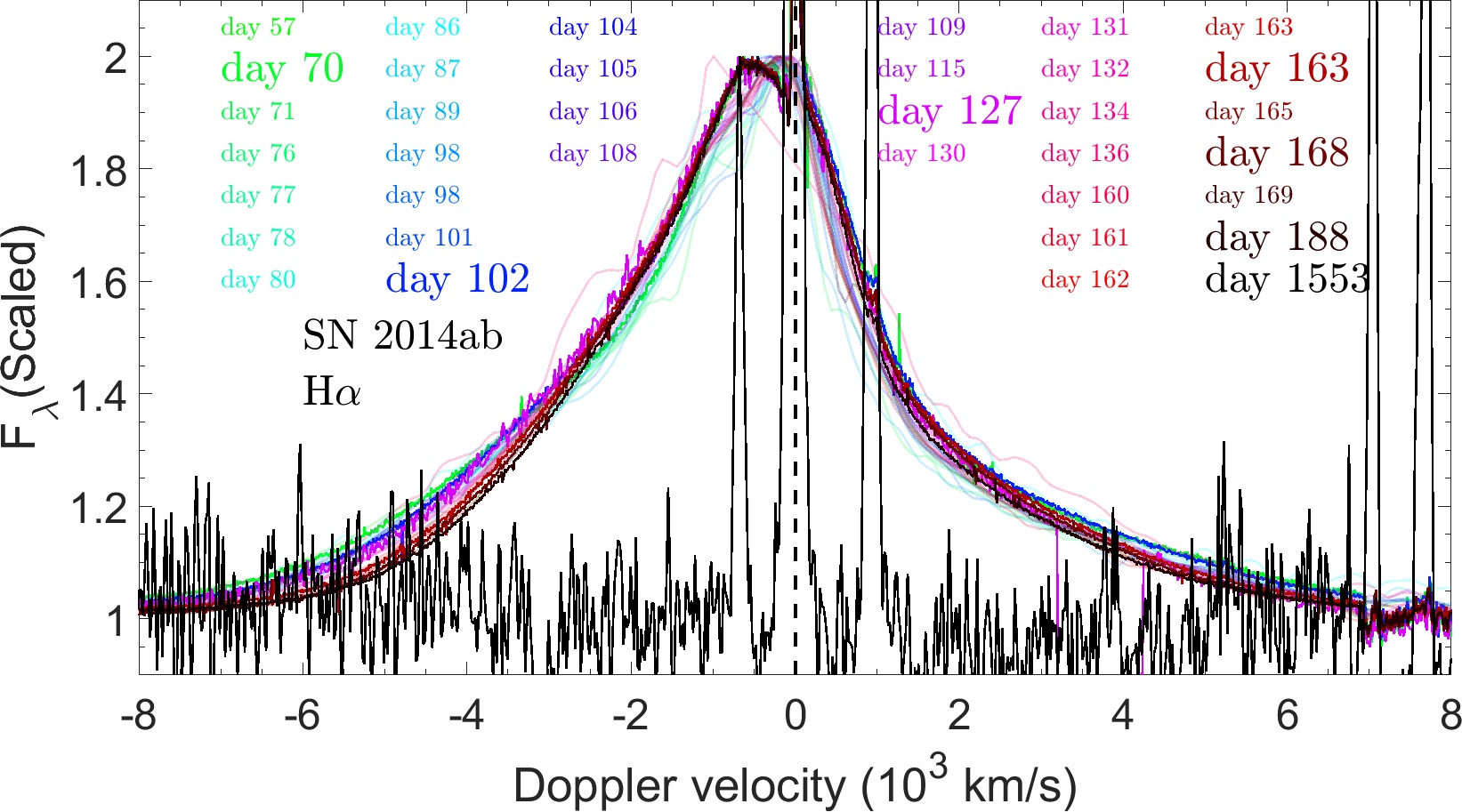}
\includegraphics[width=0.5\textwidth,clip=true,trim=0cm 0cm 0cm 0cm]{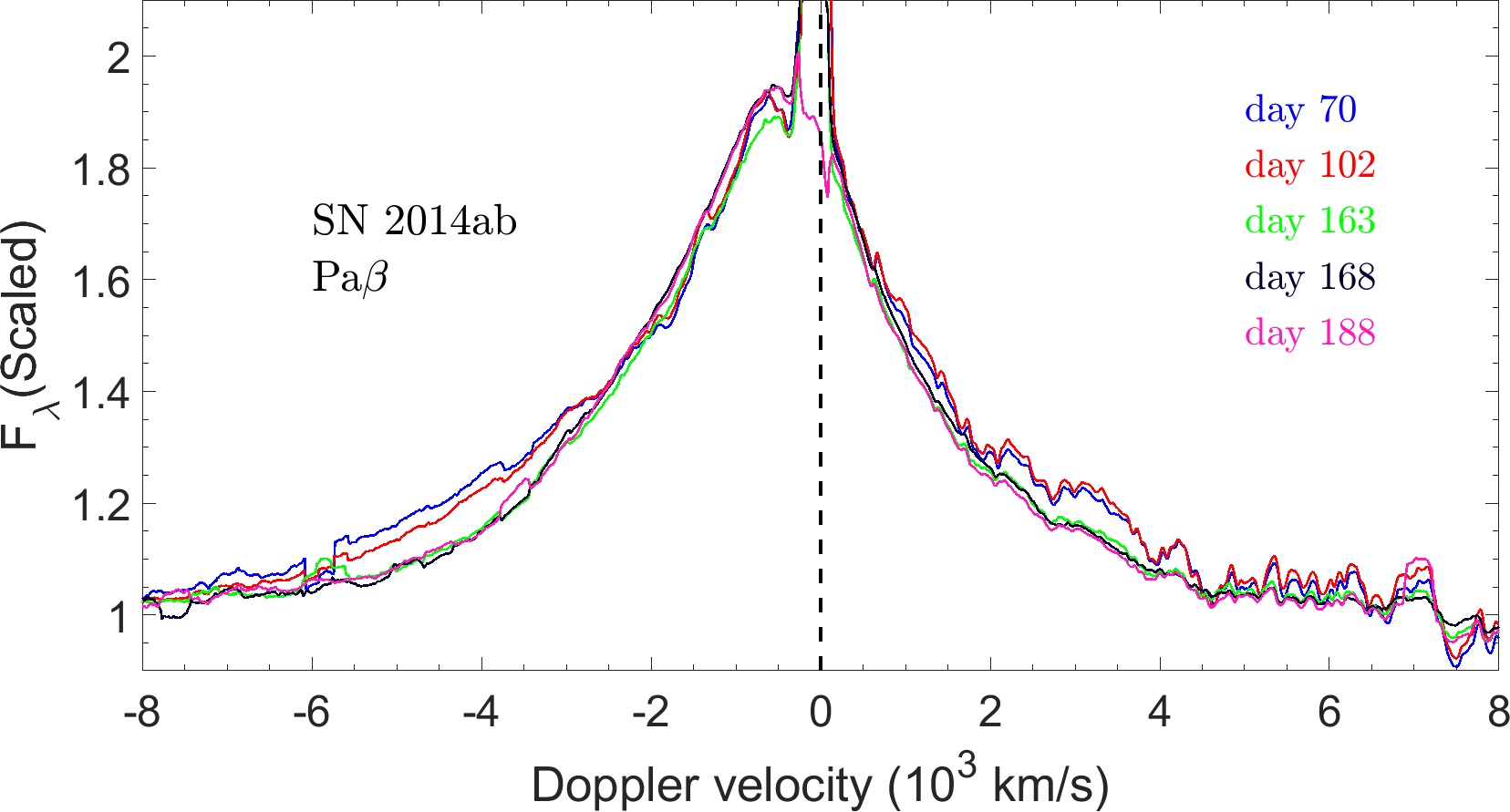}
\includegraphics[width=0.5\textwidth,clip=true,trim=0cm 0cm 0cm 0cm]{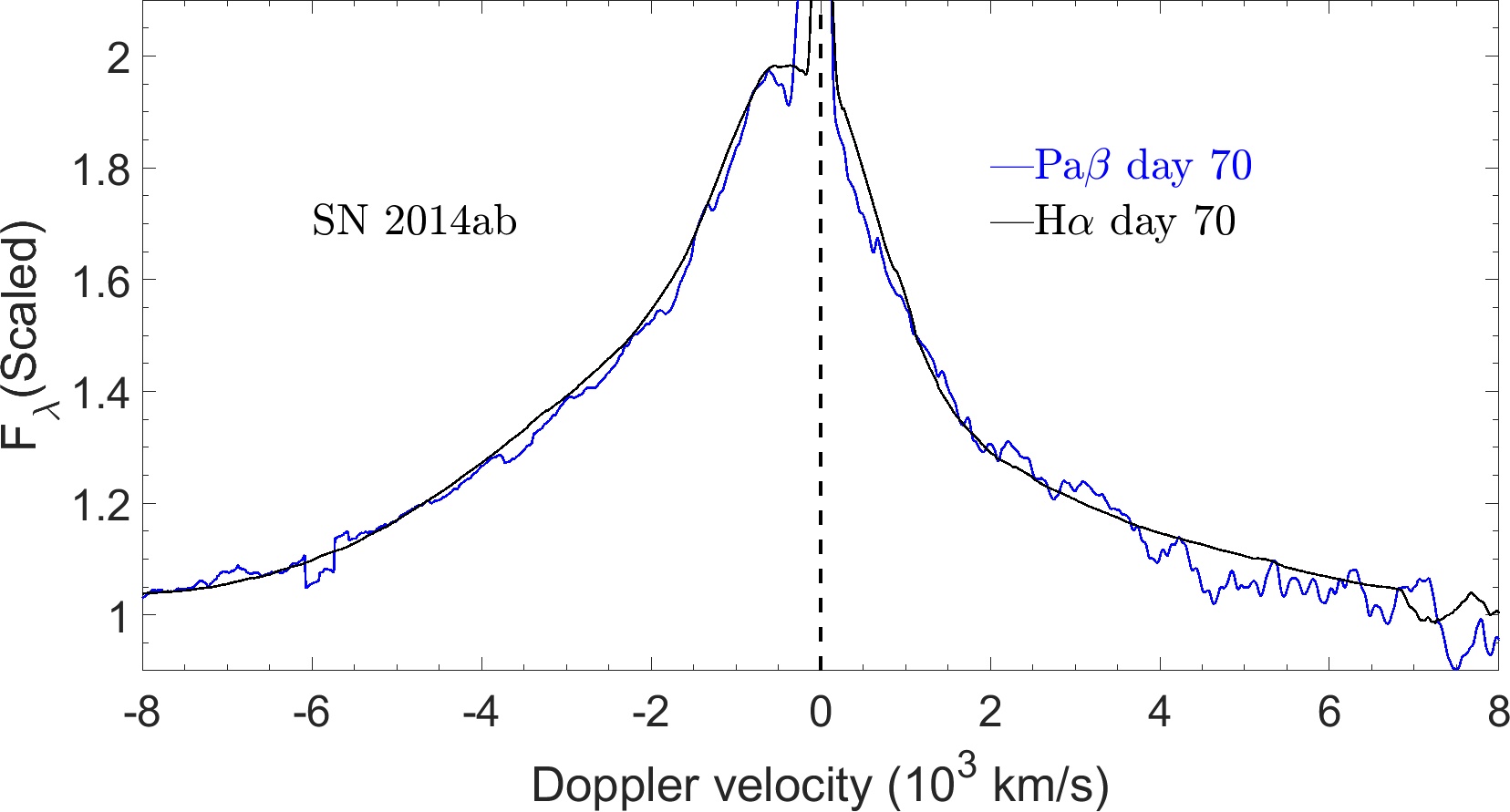}
\caption{{\it Top panel:} The progression of the H$\alpha$ line over the course of the $\sim 150$ days during which we obtained spectra.  Medium- and high- resolution spectra are shown fully opaque and have their dates labeled with larger text.  Because the spectra overlap heavily owing to a lack of change in the line profile, low-resolution spectra are shown transparently and have small date labels.  {\it Middle panel:} The progression of the Pa$\beta$ line over the course of the $\sim 70$ days during which we obtained near-IR spectra.  {\it Bottom panel:} A comparison of the first Pa$\beta$ line and its H$\alpha$ counterpart.  See \S \ref{sec:Dis:HalphavsPabeta} for a discussion of the asymmetric H$\alpha$ and Pa$\beta$ emission line profiles.}
\label{fig:HalphavsPabeta}
\end{figure}

Many of the most prominent emission features (H$\alpha$, H$\beta$, and Pa$\beta$) have intermediate-width emission components that remain constant with time throughout all of the spectra (see Fig. \ref{fig:HalphavsPabeta}).  These lines show asymmetric profiles with a net blueshift.  The blueshifted emission component of these prominent lines is stronger than the redshifted component at all epochs.  The intermediate-width H$\alpha$ emission reaches half maximum intensity at $\sim 2000$\,km\,$\mathrm{s^{-1}}$ on the blue side and $\sim 1000$\,km\,$\mathrm{s^{-1}}$ on the red side.  This intermediate-width component extends out to $\sim 4500$--6000\,km\,$\mathrm{s^{-1}}$ on the blue side before reaching 10\% of maximum intensity and $\sim 4000$--5000\,km\,$\mathrm{s^{-1}}$ on the red side before reaching 10\% of maximum intensity.  The centre of the intermediate-width component of H$\alpha$ is offset by $v \approx -500$\,km\,$\mathrm{s^{-1}}$, likely due to the CSM interaction approaching us obscuring the redshifted intermediate-width emission component.  Overall, the spectra for SN~2014ab are very similar to those of SN~2010jl, including the blueshifted profile.  However, the broad, blueshifted absorption seen in H$\beta$ in SN~2014ab is stronger than in SN~2010jl \citep{2012AJ....143...17S}.

\subsection{Spectropolarimetry}
\label{sec:Res:Specpol}
Our spectropolarimetric analysis is performed primarily using the linear Stokes parameters, $q=Q/I$ and $u=U/I$, which are rotated 45$\degree$ with respect to each other, allowing us to decompose the polarization signal into orthogonal components in position angle space. Typically, one can combine the Stokes parameters to obtain the polarization level, $P = \sqrt{Q^2 + U^2}$, and the position angle on the sky, $\theta = (1/2)\, \mathrm{tan}^{-1} (U/Q)$.  However, since the definition of the polarization makes it a positive-definite value, it may seem artificially high in cases where we have a low signal-to-noise ratio because fluctuations will raise the mean polarization level significantly.  To  partially control for this effect, we also plot the rotated Stokes parameters, which are an attempt to rotate any significant nonzero polarization signal into $qRSP$ while leaving noise in $uRSP$, though sharp changes in the $q$--$u$ signal will remain in $uRSP$ depending on the value of $\theta_{\rm smooth}$ that is used for rotation.  We use the rotated Stokes paramters $qRSP = q\, \mathrm{cos}(2\theta_{\rm smooth}) + u\, \mathrm{sin}(2\theta_{\rm smooth})$ and $uRSP = -q\,\mathrm{sin}(2\theta_{\rm smooth}) + u\, \mathrm{cos}(2\theta_{\rm smooth})$, and the optimal polarization, $P_{\rm opt} = P - \sigma_P^2 / P$ \citep{1997ApJ...476L..27W}.  The $\theta_{\rm smooth}$ value we use is a moving average over 100\,\AA, so changes in the polarization signal that occur alongside changes in the polarization angle within a small wavelength range will not be effectively rotated into $qRSP$.  

\begin{figure}
\centering
\includegraphics[width=0.5\textwidth,height=1\textheight,keepaspectratio,clip=true,trim=0cm 0cm 0cm 0cm]{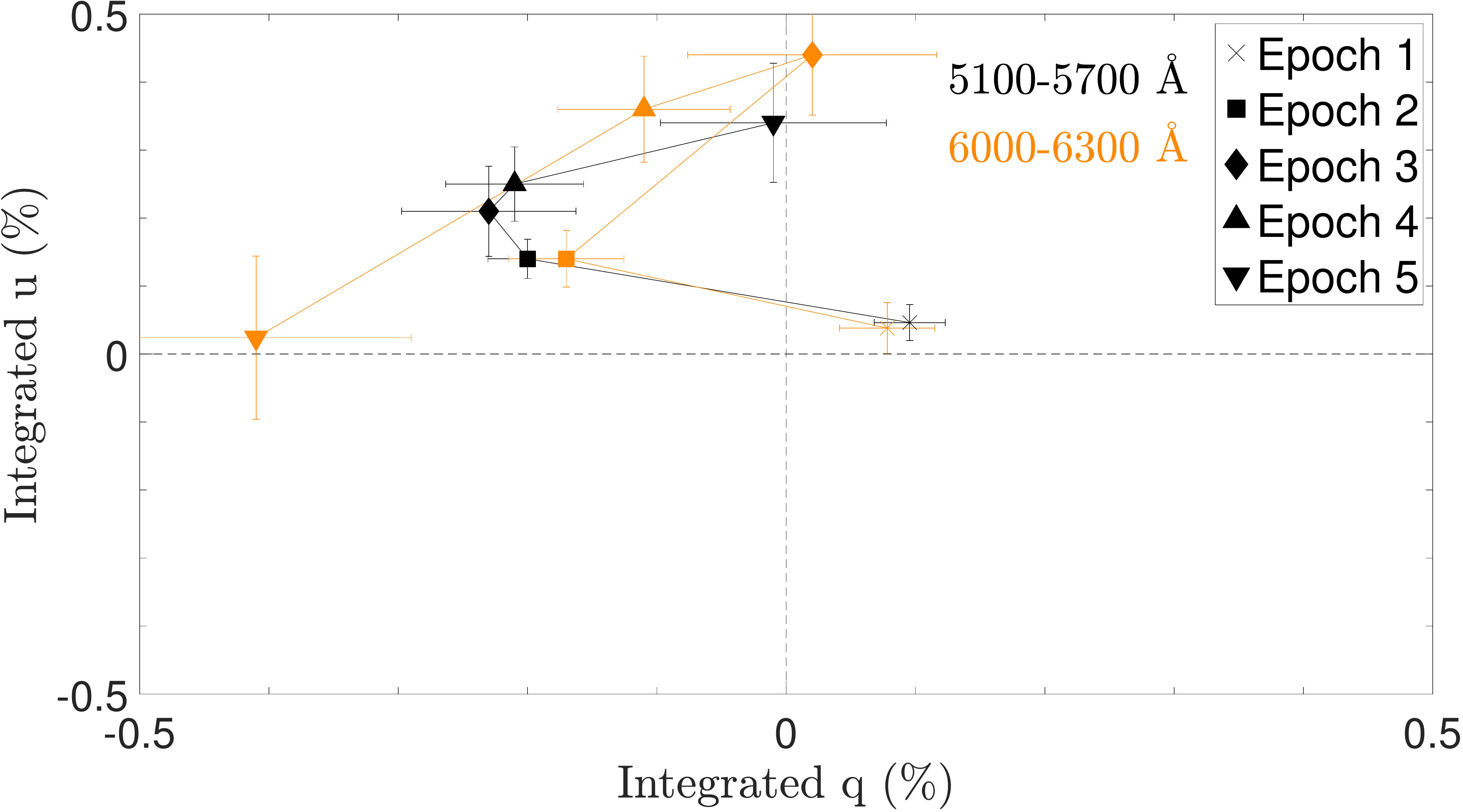}
\caption{The integrated continuum $q$ and $u$ values for the 5 different epochs of spectropolarimetry.  Error bars are estimated from the statistical rms noise and do not include systematic errors, so they are likely an underestimate of the true uncertainty.  A slight change of around 0.4--0.5\% in the intrinsic polarization of SN~2014ab is confidently shown only between epoch 1 and epochs 2-5.}
\label{fig:qucontmigration}
\end{figure}

\begin{figure*}
\centering
\includegraphics[width=1\textwidth,clip=true,trim=0cm 0cm 0cm 0cm]{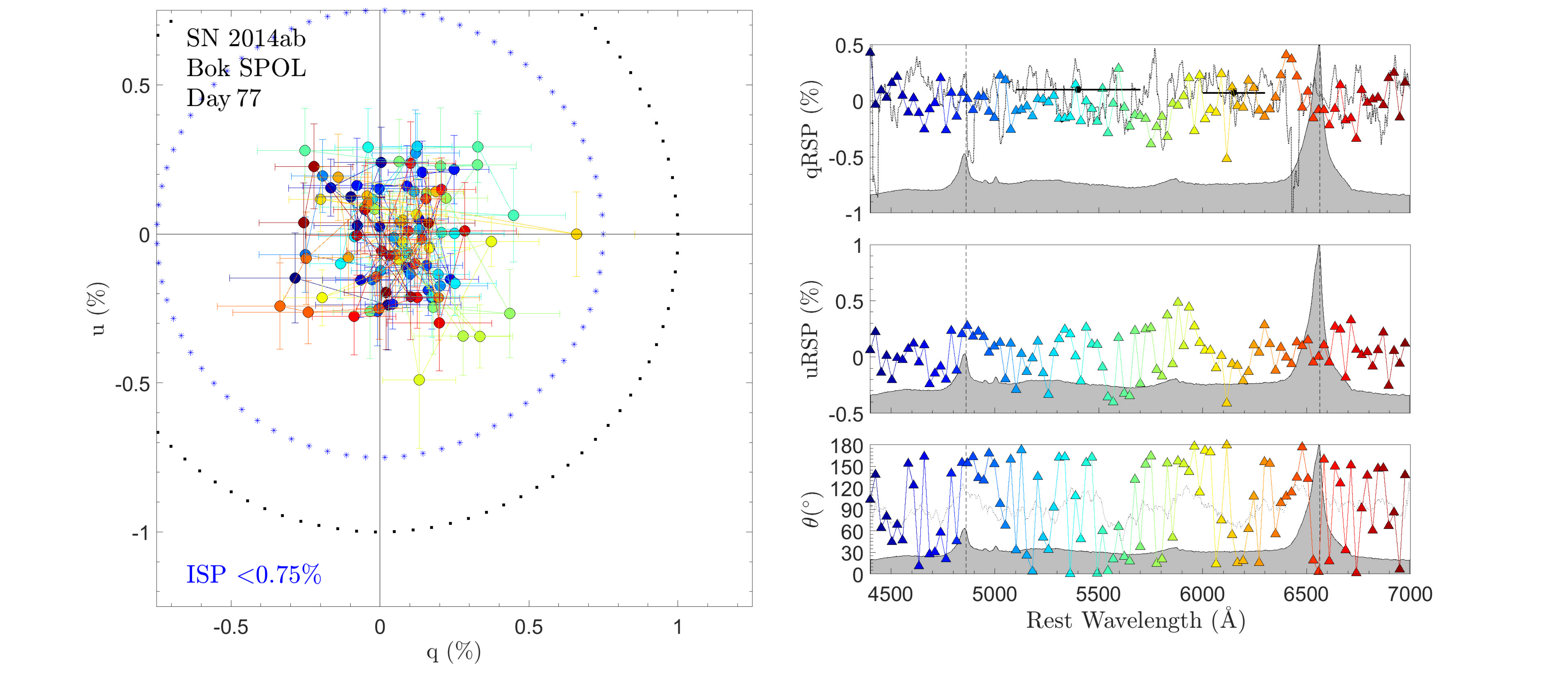}
\includegraphics[width=1\textwidth,clip=true,trim=0cm 0cm 0cm 0cm]{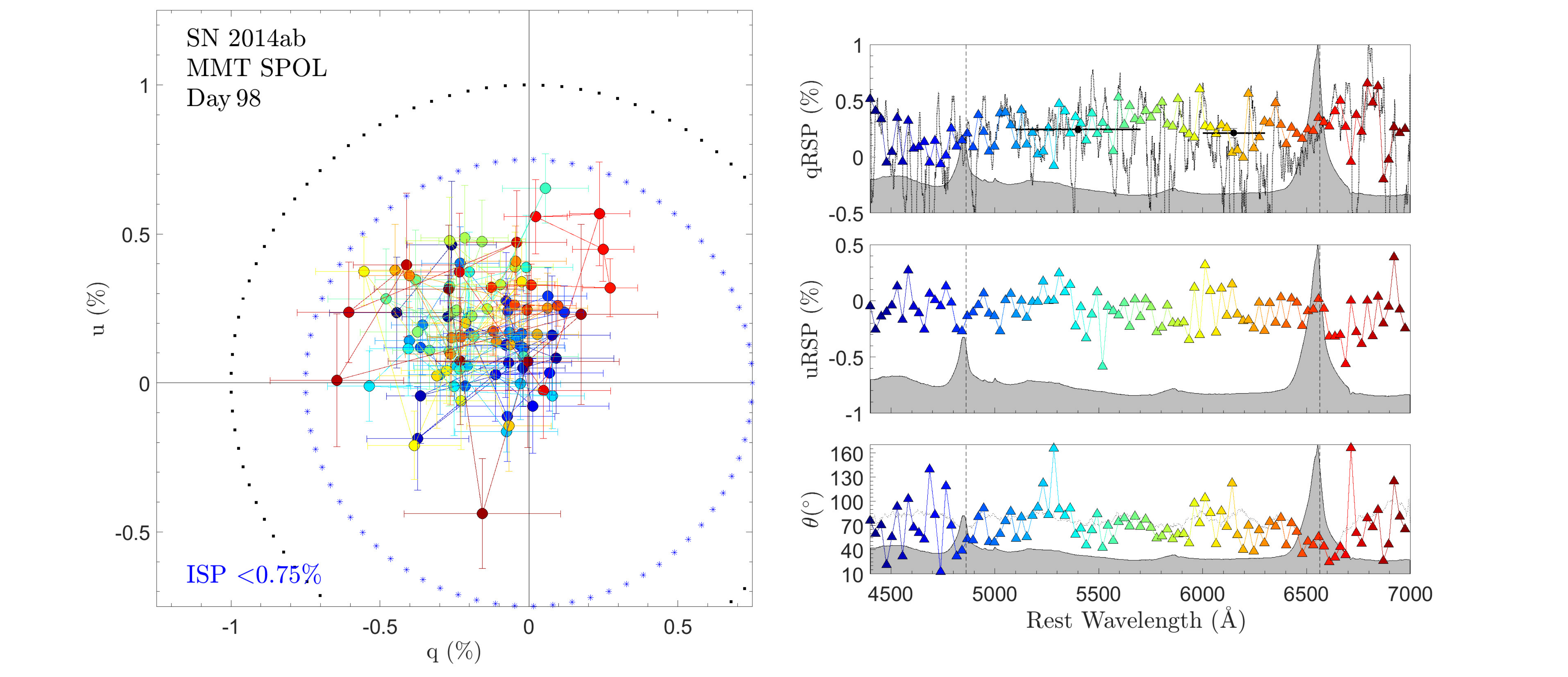}
\caption{{\it Top panels:} $q$--$u$ Stokes parameters, $qRSP$--$uRSP$ rotated Stokes parameters, and position angle $\theta$ for SN~2014ab from the 90-inch Bok telescope on days 76--80 (data from multiple days combined).  The dotted black line in the $qRSP$ plot indicates the smoothed optimal polarization value.  The dotted black line in the position-angle plot indicates the smoothed position angle, $\theta_{\rm smooth}$.  The solid black point in the $qRSP$ plot is the optimal polarization value measured across the continuum region designated by the horizontal black bars.  The data are grouped into $\sim 28$\,\AA\ bins.  Shaded regions show a scaled version of the total-flux spectrum.  Dashed lines are present at the wavelengths of H$\alpha$ and H$\beta$.  We have adopted an ISP value of $<0.75\%$ (shown as a circle of blue asterisks) based on Na~\textsc{i}~D absorption-line measurements (see \S \ref{sec:Res:Specpol}).  Black dotted circles in the $q$--$u$ plot indicate 1--2\% polarization.  Colours, bins, and error bars in the $q$--$u$ plots on the left correspond to those on the right, with wavelengths labeled on the right.  {\it Bottom panels:} The same for the MMT data from day 98.}
\label{fig:QUcircle1}
\end{figure*}

\begin{figure*}
\centering
\includegraphics[width=1\textwidth,clip=true,trim=0cm 0cm 0cm 0cm]{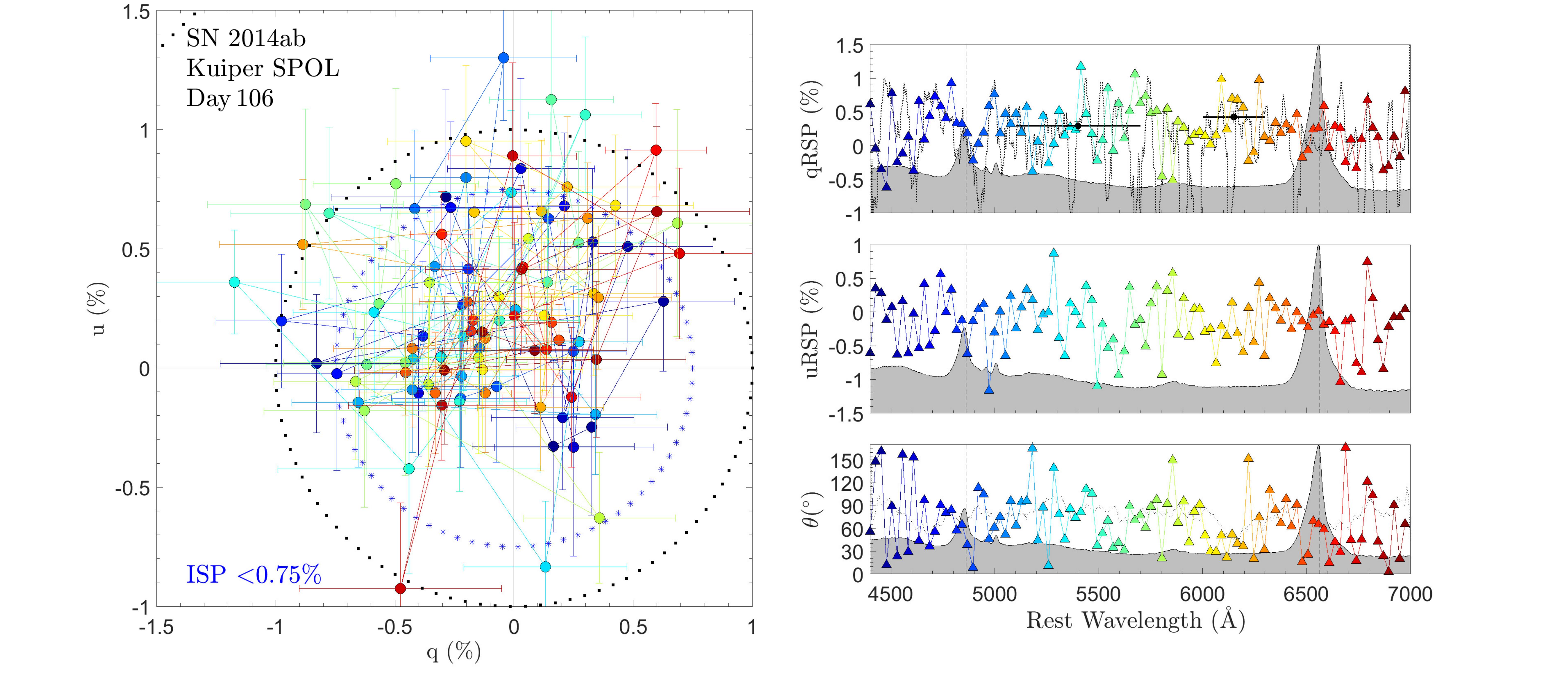}
\includegraphics[width=1\textwidth,clip=true,trim=0cm 0cm 0cm 0cm]{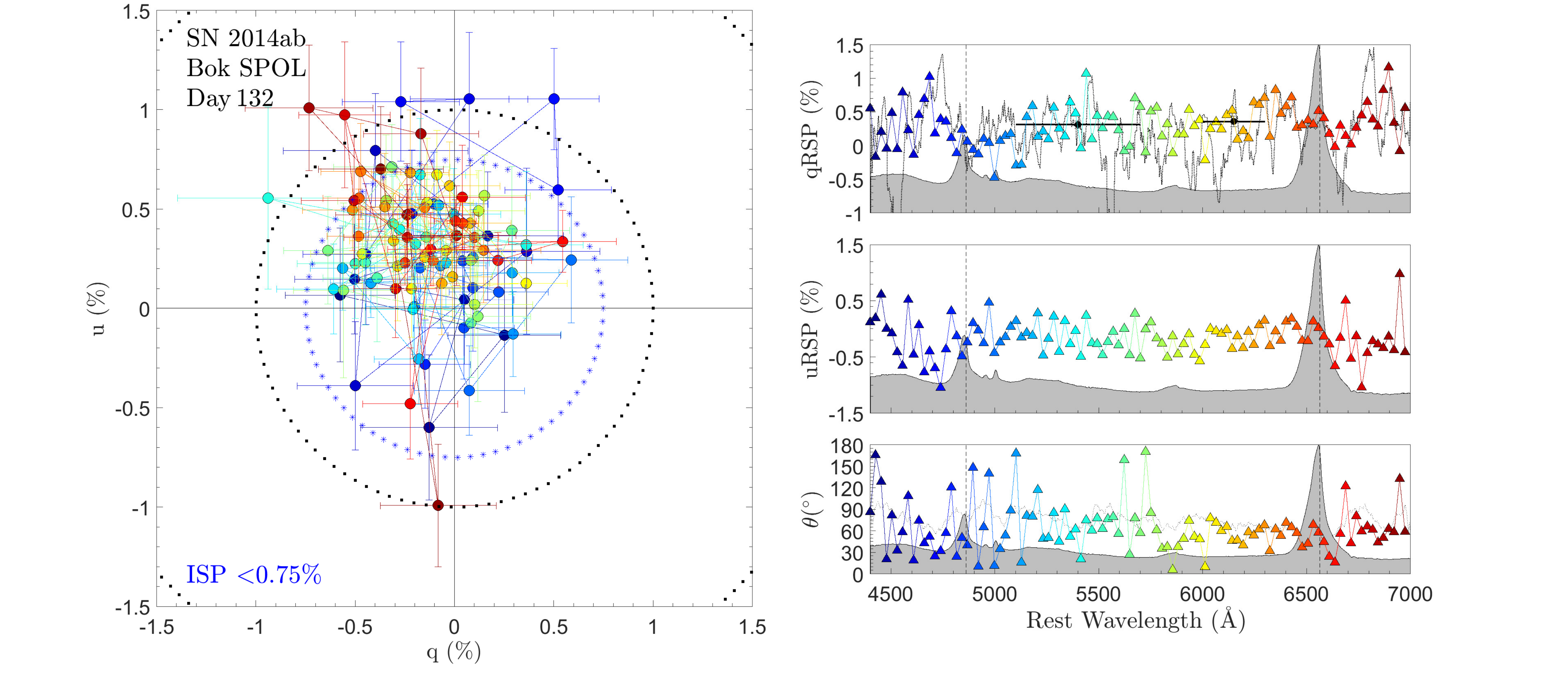}
\caption{{\it Top panels:} The same as Fig. \ref{fig:QUcircle1}, but for Kuiper data from days 104--106.  {\it Bottom panels:} The same for the 2.3\,m Bok telescope data from days 131--134.}
\label{fig:QUcircle2}
\end{figure*}

\begin{figure*}
\centering
\includegraphics[width=1\textwidth,clip=true,trim=0cm 0cm 0cm 0cm]{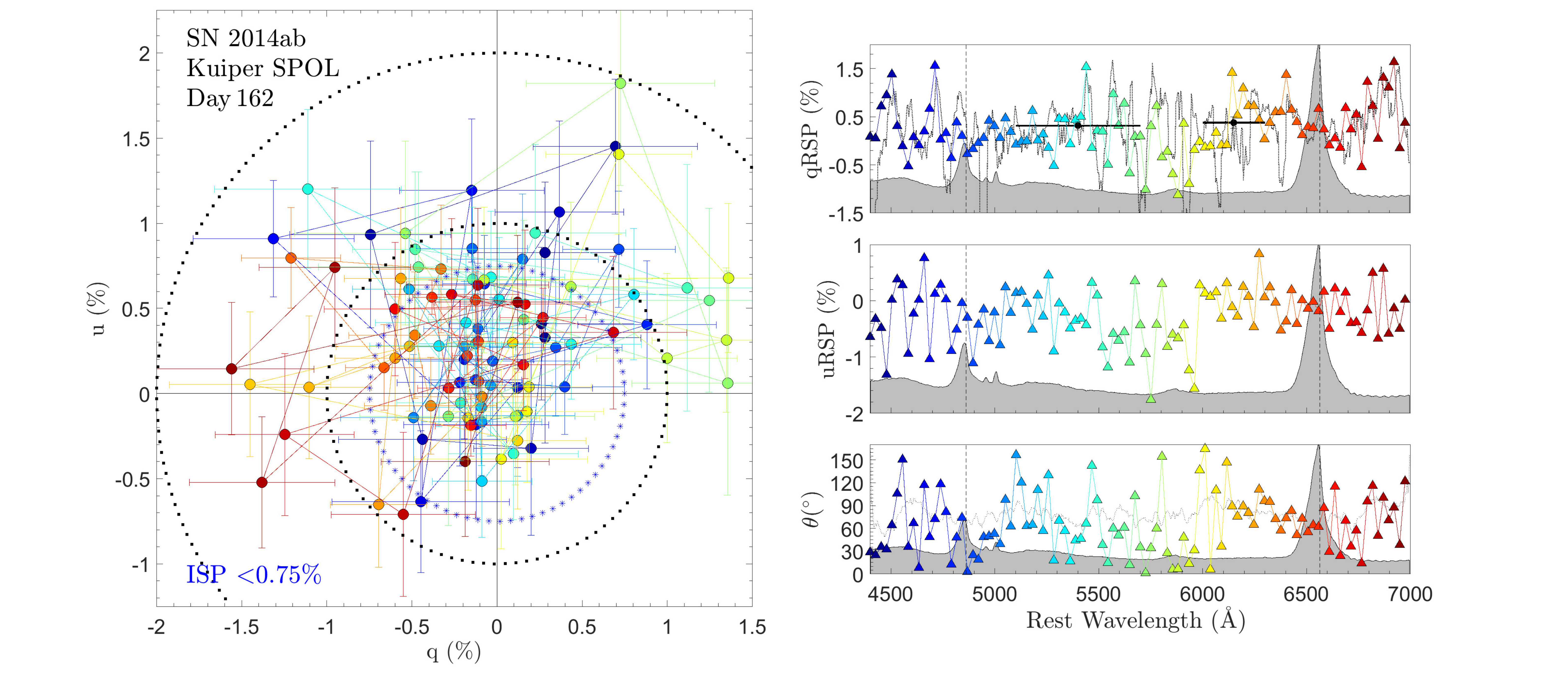}
\caption{ The same as Fig. \ref{fig:QUcircle1}, but for Kuiper data from days 161--165.}
\label{fig:QUcircle3}
\end{figure*}

We measure the optimal polarization over two wavelength ranges that are meant to avoid spectral lines in order to obtain the best measurement of the continuum polarization.  In particular, we measure the optimal polarization at 5100--5700\,\AA\ and 6000--6300\,\AA\ for each of our epochs of spectropolarimetry.  These values range between $0.07\%\pm 0.04\%$ to $0.43\%\pm 0.09\%$ (shown in Tables \ref{tab:contpolval1} and \ref{tab:contpolval2}), studied in detail in Figure \ref{fig:qucontmigration}, and are also shown in Figures \ref{fig:QUcircle1}, \ref{fig:QUcircle2}, and \ref{fig:QUcircle3}.  Uncertainty values on $q$ and $u$ were determined from the root-mean-square (rms) noise in the measured $q$ and $u$ spectra, which does not include systematic errors.  These uncertainties were then propagated to the uncertainty on the integrated $q$, integrated $u$, and optimal polarization estimates.  Although debiased, the optimal polarization level is still not a perfect measure, so we attempt to perform our analysis in the $Q$ and $U$ plane whenever possible.  This allows us to avoid problems with the positive-definite nature of polarization.

\begin{table}
\caption{Continuum Polarization Measurements Across 5100--5700 \AA}
\label{tab:contpolval1}
\begin{tabular}{cccc}
  \hline
Epoch (days) & $P_{opt}$ (\%)	&	Integrated $q$ (\%)	&	Integrated $u$ (\%)\\
	\hline
76-80	& $0.10\pm 0.03$  &    $0.10  \pm 0.03$ & $0.05\pm  0.03$\\
98	    & $0.24\pm 0.03$  &    $-0.20 \pm 0.03$ & $0.14 \pm 0.03$\\
104-109	& $0.30\pm 0.07$  &    $-0.23 \pm 0.07$ & $0.21 \pm 0.07$\\
131-134	& $0.31\pm 0.05$  &    $-0.21 \pm 0.05$ & $0.25 \pm 0.05$\\
161-165	& $0.32\pm 0.09$  &    $-0.099\pm 0.09$ & $0.34 \pm 0.09$\\
	\hline
	\end{tabular}
\end{table}

\begin{table}
\caption{Continuum Polarization Measurements Across 6100--6300 \AA}
\label{tab:contpolval2}
\begin{tabular}{cccc}
  \hline
Epoch (days) & $P_{opt}$ (\%)	&	Integrated $q$ (\%)	&	Integrated $u$ (\%)\\
	\hline
76-80	& $0.07\pm 0.04$  & $0.08  \pm 0.04$ & $0.04 \pm 0.04$\\
98	    & $0.21\pm 0.04$  & $-0.17 \pm 0.04$ & $0.14 \pm 0.04$\\
104-109	& $0.43\pm 0.09$  & $0.02  \pm 0.10$ & $0.44 \pm 0.09$\\
131-134	& $0.36\pm 0.08$  & $-0.11 \pm 0.07$ & $0.36 \pm 0.08$\\
161-165	& $0.38\pm 0.12$  & $-0.41 \pm 0.12$ & $0.02\pm 0.12$\\
	\hline
	\end{tabular}
\end{table}

We must address the tricky issue of interstellar polarization (ISP) in order to determine what level of polarization signal is actually coming from our target of interest.  It is difficult to measure the combined level of the ISP coming from the Milky Way and from the host galaxy of the SN.  Fortunately, however, Na~\textsc{i}~D absorption lines in our spectra provide constraints on the total reddening along our line of sight.  This allows us to predict very low ISP values for SN~2014ab.

In \S \ref{sec:Ext} we found a low value for the reddening of $E_{B-V} = 0.083$\,mag based on Na~\textsc{i}~D absorption plus Milky Way reddening.  We adopt this total extinction level when dereddening our spectra in Figure \ref{fig:allspectra} and all subsequent analysis.  Additionally, the ISP--reddening relation from \citealt{1975ApJ...196..261S} suggests that ISP $ < 9E_{B-V}$ \% for Milky Way dust, which means that we can use the measure of $E_{B-V}$ from the Na~\textsc{i}~D absorption lines and the Milky Way to place a constraint on the level of the ISP to $<0.75\%$.  In doing so, we have applied the relation from \citealt{1975ApJ...196..261S} to the dust in the host galaxy as well as the Milky Way, though this is not a perfect assumption \citep{2000ApJ...536..239L,2016ApJ...828...24P}.  We are not able to constrain the likely location of the total ISP in the $q$--$u$ plane, so we do not subtract the ISP directly from our measurements.  If the full extent of the ISP were aligned exactly opposite in the $q$--$u$ plane to our strongest contiuum polarization signal measurement for SN~2014ab ($0.43\%\pm 0.04\%$), then SN~2014ab would have a continuum polarization and line polarization of 1.18\%.  We use this as a conservative upper limit on the continuum and line polarization for SN~2014ab.  This is lower than that for other SNe~IIn with published spectropolarimetry (SN~1997eg: \citealt{2008ApJ...688.1186H}; SN~1998S: \citealt{2000ApJ...536..239L}; SN~2006tf: \citealt{2008ApJ...686..467S}; SN~2009ip: \citealt{2014MNRAS.442.1166M,2017arXiv170108885R}; SN~2012ab: \citealt{2018MNRAS.475.1104B}) and is discussed in more detail \S \ref{sec:Dis:CircSym}.

Figures \ref{fig:QUcircle1}, \ref{fig:QUcircle2}, and \ref{fig:QUcircle3} show the spectropolarimetric data plotted in the $q$--$u$ plane.  The polarization signal is weak at every one of our 5 epochs spanning from day 76 to day 165, although the signal becomes much noisier by our later epochs.  The $q$ and $u$ values are relatively evenly scattered around the origin in every epoch, though we see a slight shift in the optimal polarization value across the continuum between the first epoch of spectropolarimetry and the remaining four epochs.  Figure \ref{fig:qucontmigration} shows the $q$ and $u$ values integrated across two wide continuum regions (5100--5700\,\AA\ and 6100--6400\,\AA).  The error bars are likely an underestimate of the true uncertainty, so we conclude that only the shift in polarization from the first epoch to the later epochs is statistically significant and consistent across both continuum wavelength regions.  

\section{Discussion}

\subsection{SN, AGN, or TDE}
\begin{figure}
\centering
\includegraphics[width=0.5\textwidth,height=0.5\textheight,keepaspectratio,clip=true,trim=0cm 0cm 0cm 0cm]{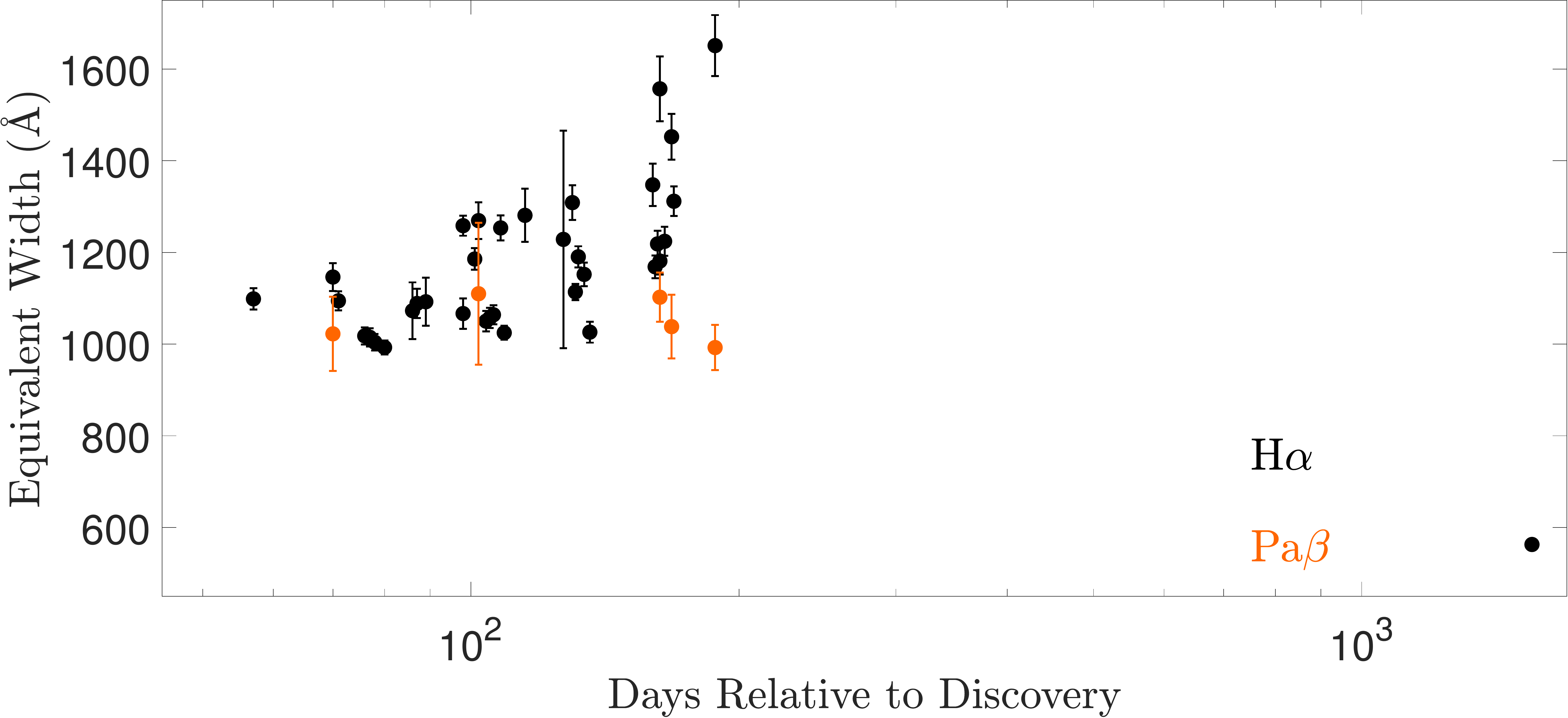}
\caption{The equivalent width of H$\alpha$ (black) and Pa$\beta$ (orange) at all epochs.  As the light curve of SN~2014ab fades $\sim 1$\,mag over the course of $\sim 150$ days (so that the visual continuum fades by a factor of 2--2.5), the equivalent width of H$\alpha$ nearly doubles.}
\label{fig:EQW}
\end{figure}

Because SN~2014ab is within 1$^{\prime\prime}$ of its host-galaxy centre, we consider the possibility that it is an active galactic nucleus (AGN) or tidal disruption event (TDE).  The light curve of SN~2014ab (see \S \ref{sec:Res:LC}) shows a relatively smooth drop of $\sim 1$\,mag over the course of $\sim 150$ days.  During this period of time, the equivalent width of the H$\alpha$ line increases roughly twofold (see Figure \ref{fig:EQW}), probably caused by a fading continuum paired with a persistent bright CSM interaction region.  While AGNs can exhibit such variations on this timescale \citep{1997ARA&A..35..445U}, they do not generally brighten and fade only once over the course of a $\sim 11$\,yr timescale.  Additionally, AGNs tend to have a number of forbidden emission lines which we do not detect.  These include [O~I] $\lambda\lambda$6300, 6364, [O~III] $\lambda\lambda$4959, 5007, [N~I] $\lambda$5199, and [N~II] $\lambda\lambda$6548, 6583.  The He~II $\lambda$4686 line is often quite strong compared to the Balmer series in AGNs, but we do not detect it in our spectra.  The late-time spectrum of SN~2014ab taken on day 1553 shows only narrow H$\alpha$ emission, whereas the intermediate-width component is completely gone.  For these reasons, we find it unlikely that SN~2014ab is an AGN.


We also consider the possibility that SN~2014ab may be a TDE because TDEs also produce bright asymmetric signatures and are found coincident with the central region of their host galaxy.  Unlike an AGN, a TDE is consistent with the long stable CSS photometry for years prior to the brightening event, followed by a decrease in brightness that does not rebrighten again.  We cannot easily compare our light curve to the theoretical decline rate in TDEs ($t^{-5/3}$; \citealt{1988Natur.333..523R,1989ApJ...346L..13E,1989IAUS..136..543P}) because we do not know when SN~2014ab reached its peak brightness.  However, our light curve shows a steepening in the decline of SN~2014ab and a duration of over 150 days, which is uncharacteristic of TDEs.  If the event we observed was in fact a TDE, we would expect the light curve to become gradually flatter.  Spectroscopically, we would expect there to be an offset between the narrow lines and the intermediate-width lines in the case of a TDE, which we do see \citep{2009MNRAS.400.2070S}.  However, we do not detect any strong He~II emission, which is thought to be common in TDEs, though not always detected \citep{2012Natur.485..217G,2014ApJ...793...38A}.  Overall, the shape of the light curve, the long duration of the event, and the lack of He~II emission suggest that SN~2014ab is not a TDE.  Henceforth, we assume that SN~2014ab is a core-collapse SN with strong CSM interaction.

\subsection{Light Curve}
\label{sec:Dis:LC}
SN~2014ab is a luminous example among SNe~IIn.  Its projected peak in the Nickel photometry was at $M_V =-19.14$\,mag, approaching values of superluminous SNe, and it remained bright, declining slowly  over the next 211 days to a brightness of $-17.89$\,mag (see \S \ref{sec:Res:LC} for a summary of the photometric data).  

\begin{figure*}
\centering
\includegraphics[width=1\textwidth,height=1\textheight,keepaspectratio,clip=true,trim=0cm 0cm 0cm 0cm]{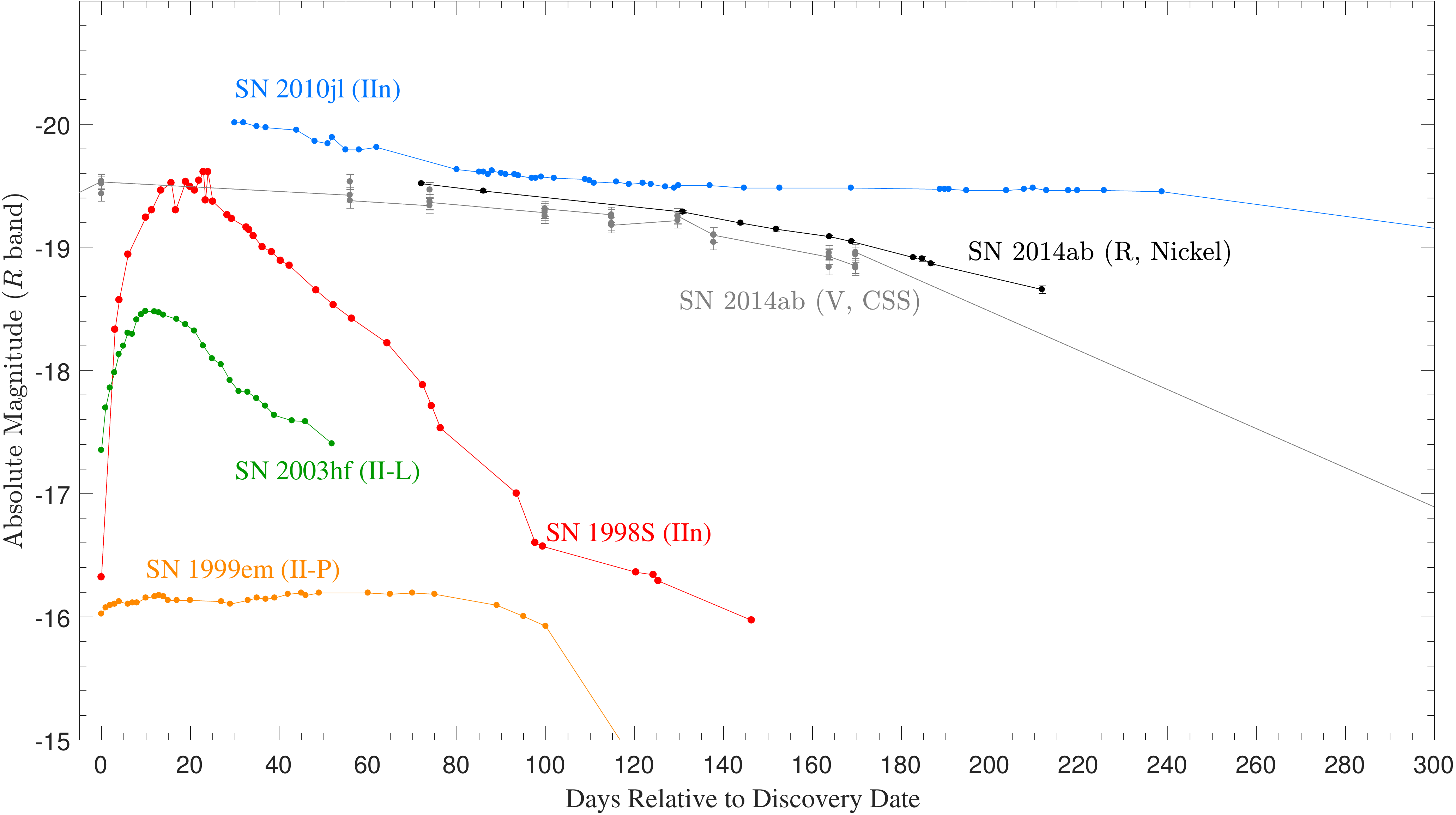}
\caption{$R$-band (Nickel telescope) and $V$-band (CSS, host galaxy subtracted) observations of SN~2014ab relative to the discovery date of 2014 Jan. 12.4755.  For comparison, we have included the $R$-band light curves of the Type II-P SN~1999em (orange; \citealp{2002PASP..114...35L}), the Type II-L SN~2003hf (green; \citealp{2014MNRAS.445..554F}), the Type IIn SN~1998S (red; \citealp{2000MNRAS.318.1093F,2011arXiv1109.0899P}), and the Type IIn SN~2010jl (blue; G. Williams, private communication).}
\label{fig:LCcompare}
\end{figure*}

The sustained high luminosity suggests that SN~2014ab is more akin to superluminous SNe~IIn like SN~2010jl \citep{2011A&A...527L...6P} and SN~2006tf \citep{2008ApJ...686..467S}, albeit with a slightly lower peak luminosity, and quite different from faster declining SNe~IIn like SN~1998S \citep{2000MNRAS.318.1093F}.  Of course, the true peak luminosity of SN~2014ab may have been higher than its observed peak because of its relatively late discovery.  Figure \ref{fig:LCcompare} shows a comparison of several SN light curves.  Although some SNe~IIn (such as SN~1998S) exhibit light-curve decline rates similar to those of SNe~II-L (such as SN~2003hf), other SNe~IIn (such as SN~2014ab and SN~2010jl) exhibit slower light-curve decline rates for much longer durations.  We attribute the extended and bright plateau in SN~2014ab to strong CSM interaction.

\subsection{Lack of Spectral Evolution}
\label{sec:Dis:LackofSE}
Although SN~2014ab drops in brightness significantly over the course of the $\sim 150$ days during which we also have spectra, the spectra show little qualitative change.  This suggests that we did not observe SN~2014ab during its early stages, as most SNe~IIn  are observed to have drastically changing early-time spectra as they transition from optically thick CSM and electron scattering line profiles to later times when emission from the cold dense shell is revealed \citep{smith17}.  Instead, we are likely observing SN~2014ab while the SN ejecta are crashing into increasingly distant but smoothly distributed CSM, creating spectral features that are similar at each epoch because the advancing shock speed and CSM density change slowly.

\subsection{CSM Interaction Luminosity}
The strong intermediate-width H$\alpha$ component seen in our spectra and the spectral similarity to SN~2010jl suggest that CSM interaction dominates the emitted radiation in SN~2014ab.  If the CSM interaction is powering most of the luminosity, then we can estimate a lower limit to the wind-density parameter ($w = \dot{M}_{\rm CSM}/v_{\rm w}$, where $v_{\rm w}$ is the velocity of the preshock wind) prior to the explosion of SN~2014ab. We calculate the wind-density parameter as 

\begin{equation}
w = 2L/v_{\rm SN}^3,
\end{equation}

\noindent where $L$ is the observed luminosity and $v_{\rm SN}$ is the velocity of the postshock shell \citep{2008ApJ...686..467S}.

The velocity of the postshock shell is determined from the intermediate-width component of the H$\alpha$ line, which is measured to have a half width at half-maximum intensity of $\sim 2000$\,km\,s$^{-1}$.  Although our $R$-band measurements do not extend to day 0, we assume a rough average $M_{R} \approx -19$\,mag over the course of 200 days, and conservatively adopt no bolometric correction, to estimate the average luminosity ($L \approx 3.1 \times 10^{9}\,{\rm L}_{\odot}$).  We obtain a conservative estimate of the wind-density parameter of $\sim 3 \times 10^{18}$\,g\,$\mathrm{cm^{-1}}$.  If we assume a steady wind velocity of $\sim 80$\,km\,s$^{-1}$ (estimated from the P-Cygni absoprtion in H$\alpha$) from the progenitor, we can estimate the mass-loss rate to be at least $\dot{M} \approx \,{\rm M}_{\odot}\,\mathrm{yr^{-1}} (v_{\rm w}/(80\,\mathrm{km}\,\mathrm{s^{-1}}))0.4$. This mass-loss rate is higher than any normal steady wind mass loss, but is achievable by episodic super-Eddington winds of massive stars \citep{so06}. The intermediate-width component of H$\alpha$ persists for over 150 days (and is likely a significant source of the luminosity of SN~2014ab going back another 50 days), so we can estimate the duration of its  pre-SN mass-loss episode by determining how long pre-SN mass loss must have been occurring.  In order for the $\sim 2000$\,km\,s$^{-1}$ shell to be continually running into CSM which was ejected at roughly $80$\,km\,s$^{-1}$ for 200 days, the pre-SN mass loss must have begun at least $T\, v_{\rm SN}/v_{\rm w} = 14$\,yr before explosion, where $T$ is the duration of the strong intermediate-width component of H$\alpha$.  This suggests a total mass loss of $\sim 5\,{\rm M}_{\odot}$ in the 14\,yr prior to explosion, comparable to the CSM mass inferred for some of the most luminous SNe~IIn \citep{smith07,sm07,2014ApJ...781...42O}.

The variety of progenitor candidates for SNe~IIn have widely differing mass-loss rates.  \citet{2014ARA&A..52..487S} estimated that eruptive luminous blue variable (LBV) progenitors can have mass-loss rates in the range of 0.01--10\,M$_{\odot}\,\mathrm{yr^{-1}}$, while even very massive red supergiant or yellow hypergiant progenitors have mass-loss rates in the range of $10^{-4}$--$10^{-3}$\,M$_{\odot}\,\mathrm{yr^{-1}}$.  LBVs undergoing eruption are therefore the only known class of progenitors to SNe~IIn that have mass-loss rates as high as those we measure for SN~2014ab \citep{so06,2014ARA&A..52..487S}. The physical cause of this eruptive mass loss remains unknown, but several ideas have been proposed, including wave-driven mass loss during Ne, O, or Si burning \citep{2012MNRAS.423L..92Q}, pulsational pair eruptions \citep{2017ApJ...836..244W} or other nuclear burning instabilities \citep{2014ApJ...785...82S}, and various types of binary interaction \citep{2012ApJ...752L...2C,2014ApJ...785...82S}.

\subsection{Asymmetric H$\alpha$ and Pa$\beta$}
\label{sec:Dis:HalphavsPabeta}
\begin{figure}
\centering
\includegraphics[width=0.5\textwidth,height=0.5\textheight,keepaspectratio,clip=true,trim=0cm 0cm 0cm 0cm]{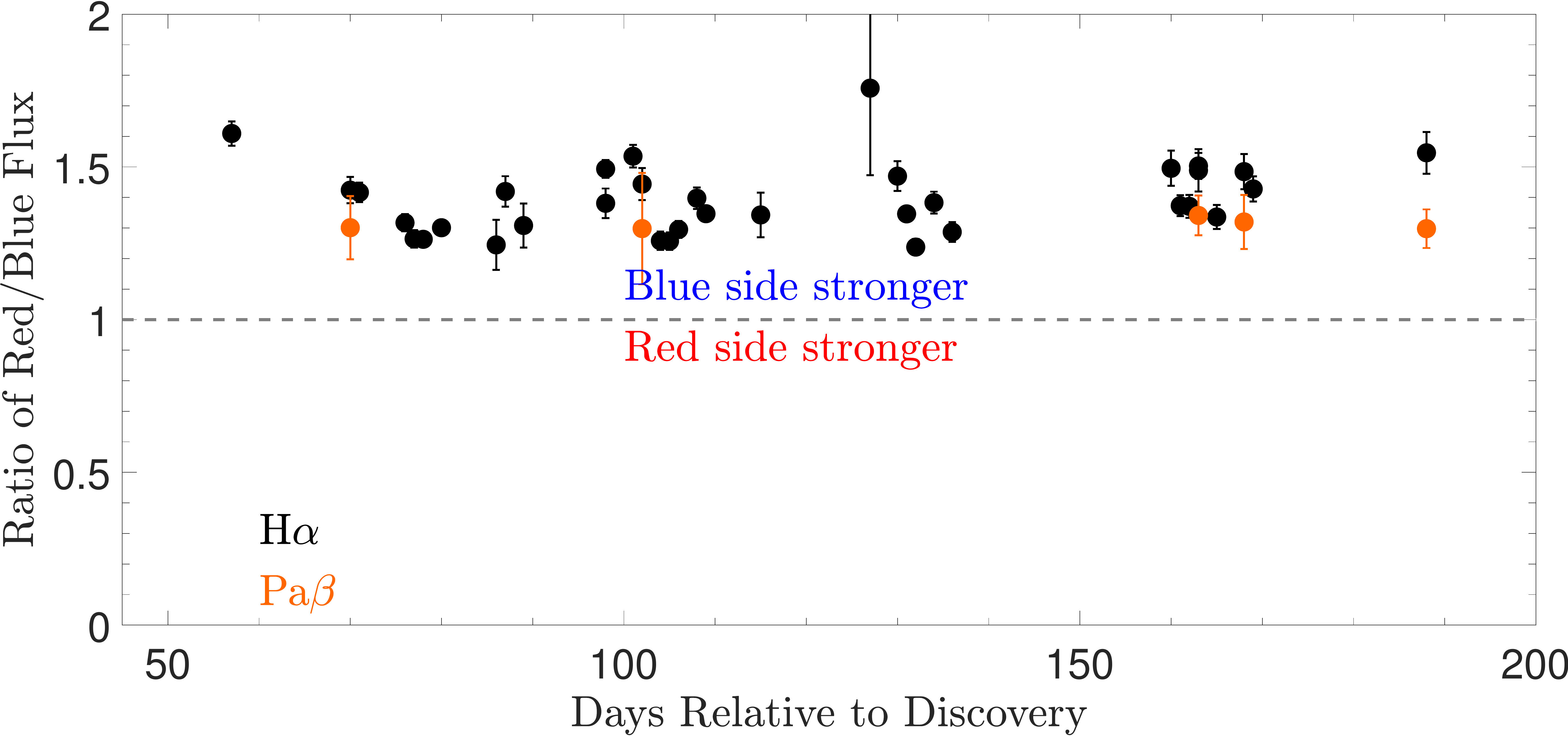}
\caption{Ratio of the flux for the blueshifted side of a line compared to the redshifted side of a line, for H$\alpha$ (black) and Pa$\beta$ (orange).  This gives a quantitative indication of the relative asymmetry in the line profile.  Note that the statistical errors shown for Pa$\beta$ are likely an underestimate of the true error because the continuum level on the redshifted side of the line is extremely uncertain owing to blending with telluric absorption.}
\label{fig:Halphabvsrratio}
\end{figure}

Figure \ref{fig:HalphavsPabeta} shows the H$\alpha$ and Pa$\beta$ emission-line profiles, scaled to the continuum fit beyond the wings of each line.  As the red wing of Pa$\beta$ is blurred with telluric features, this continuum estimate contains greater uncertainty not reflected in the line-strength measurements shown in Figure \ref{fig:Halphabvsrratio}.  The blueshifted side of both lines is clearly stronger than the redshifted side of the line at all epochs.  The early-time spectra for both emission lines are slightly wider on the blueshifted side of the line than in the late-time spectra.  The H$\alpha$ emission line in the day 57 spectrum has a width on the blueshifted side at 10\% of maximum intensity of $\sim 6000$\,km\,s$^{-1}$, whereas the day 188 spectrum has a width on the blueshifted side at 10\% of maximum intensity of $\sim 4500$\,km\,s$^{-1}$.  Emission in the wings of the line likely arises in the SN ejecta, with slower SN ejecta hitting the shock front as time progresses.  It is possible we are not seeing this same effect on the redshifted side of the emission line because it is obscured by the blueshifted CSM interaction region (see \S \ref{sec:Dis:PhysPic} and Figure \ref{fig:SN2014abCartoon} for a more detailed discussion of the line of sight).

We measure the flux of the H$\alpha$ line above the continuum level on the redshifted and blueshifted sides of the line separately ($v = 0$ is chosen to align with the narrow component in high-resolution spectra and where we would expect the narrow component in low-resolution spectra based on temporal interpolation between the high-resolution spectra).  The blue/red ratio of these  fluxes is shown in Figure \ref{fig:Halphabvsrratio}.  Throughout all epochs, we find that the blueshifted component of the H$\alpha$ line is roughly 1.4 times as strong as the redshifted component.  This may in principle be due to dust formation, occultation by the SN ejecta or CSM interaction region, electron scattering, or real geometrical asymmetries.  Since the SN was first observed during the decline phase in its light curve and the spectra remain surprisingly similar throughout all epochs even as the SN fades, we find it unlikely that the SN ejecta are still optically thick in the continuum at late times.  If the SN ejecta were optically thick initially, then over the course of the 150 days during which we obtain spectra, we would have expected the redshifted component of the spectra to become relatively stronger as the SN ejecta become less optically thick.  

\begin{figure}
\centering
\includegraphics[width=0.5\textwidth,height=1\textheight,keepaspectratio,clip=true,trim=0cm 0cm 0cm 0cm]{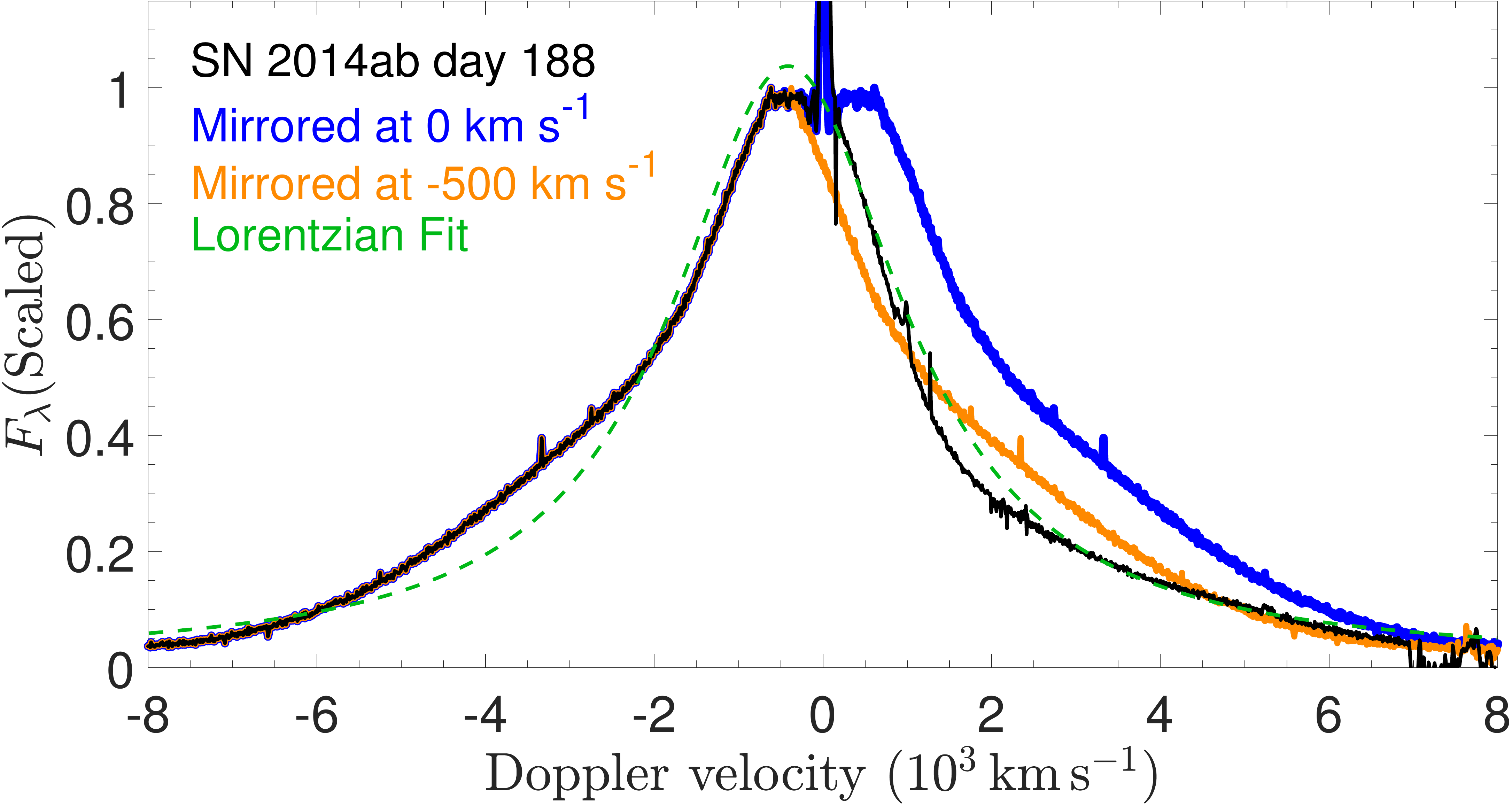}
\caption{The day 188 ESO-VLT X-shooter spectrum of H$\alpha$ is shown in black.  The blueshifted wing of the line has also been mirrored across a vertical line at 0\,km\,s$^{-1}$ (blue) and at $-500$\,km\,s$^{-1}$ (orange).  A best-fit Lorentzian is overplotted as a dashed green line.  A Gaussian, Lorentzian, and Voigt profile all fail to fit the asymmetric wings of the H$\alpha$ emission-line profile.}
\label{fig:MirAsym}
\end{figure}

In order to probe whether dust formation is likely, we also perform the same blueshifted vs. redshifted component measurement on the Pa$\beta$ line, though our estimates of the redshifted flux are prone to noise from atmospheric absorption.  The blue/red ratio of fluxes for Pa$\beta$ is shown in Figure \ref{fig:Halphabvsrratio}.  Dust extinction by small grains is wavelength dependent as was found in SN~2010jl \citep{2012AJ....143...17S}, but we do not see a wavelength-dependent asymmetry in the lines of H$\alpha$ and Pa$\beta$ in SN~2014ab.  Instead, we see a similar 1.4:1 strength in the blueshifted component compared to the redshifted component for both H$\alpha$ and Pa$\beta$.  Lastly, electron scattering has been shown to produce augmented blueshifted emission profiles that are offset from the rest frame given high velocities \citep{2009MNRAS.394...21D}.  However, the H$\alpha$ and Pa$\beta$ emission-line profiles we observe have wider blue wings than red wings (see Figure \ref{fig:MirAsym}), unlike those seen in high-velocity electron scattering profiles.  We also do not observe strong blueshifted absorption that is suggested in the case of electron scattering.  This implies that occultation by the optically thick CSM interaction region, extinction  from very large dust grains, or real geometrical asymmetries in the SN ejecta or CSM are the cause of these asymmetric line profiles.

\subsection{Weak Symmetry in the Face-on Viewing Plane}
\label{sec:Dis:CircSym}
We estimate a low value for the ISP ($<0.75\%$) based on low Na~\textsc{i}~D equivalent widths and the relation between these equivalent widths and reddening provided by \citet{2012MNRAS.426.1465P}.  The $q$ and $u$ values for the first four epochs of spectropolarimetry fall mostly within the constraint on the ISP (see Fig. \ref{fig:QUcircle1} and \ref{fig:QUcircle2}).  The $q$ and $u$ values for the last epoch of spectropolarimetry extend to about twice the estimate of the ISP and have much larger error bars, but still average a polarization measurement below that of the ISP constraint (see Fig. \ref{fig:QUcircle3}).  Figure \ref{fig:qucontmigration} shows a slight change in the integrated $q$ and $u$ values between the first epoch and the last four epochs of spectropolarimetry.  This is also seen by the slight increase in the optimal polarization measured in the continuum shown in Tables \ref{tab:contpolval1} and \ref{tab:contpolval2}.  The change in the continuum polarization cannot be caused by the relatively constant ISP, and therefore we conclude that there is a slight change ($0.36\%$) in the intrinsic polarization of SN~2014ab.  This slight change sets a lower limit to the instrinsic polarization to SN~2014ab of at least $0.36\%$, which when combined with our upper limit of $1.18\%$ discussed in \S \ref{sec:Res:Specpol} suggests that SN~2014ab has low continuum polarization compared to other SNe~IIn with published spectropolarimetry.  Thus, despite the asymmetries measured in the emission lines of SN~2014ab, spectropolarimetry at five different epochs (spanning from day 76 to day 165) suggests that the face-on viewing plane has much less deviation from circular symmetry compared to other SNe~IIn.

\subsection{Comparison to SN~2010jl}
SN~2014ab shows many similarities to SN~2010jl, particularly in its near-IR spectra.  Near-IR spectra of both SN~2014ab and SN~2010jl are plotted in Figure \ref{fig:IRspectra} for comparison.  The major features (He~\textsc{i} $\lambda$10,830, Pa$\beta$, Br$\delta$, and Br$\gamma$) are all very similar in both objects.  Of particular interest is the fact that both Pa$\beta$ lines have augmented intermediate-width blueshifted emission offset from the centre of the narrow Pa$\beta$ emission component.  Unlike SN~2014ab, SN~2010jl was observed to have a symmetric line profile for H$\alpha$ at early times (days 29 and 59, \citealt{2012AJ....143...17S,2014ApJ...797..118F}), which then evolved to the blueshifted profile similar to SN~2014ab by day 85.  Of course, SN~2014ab may have also had symmetric lines in its early evolution, since it seems to have been discovered relatively late.  At late times (day 448 onward; \citealt{2014ApJ...797..118F}), SN~2010jl is seen to be dominated by thermal emission from dust.  When comparing the H$\alpha$ emission-line profile to that of Pa$\beta$, \citet{2012AJ....143...17S} found that the blueshifted dominance is more pronounced at shorter wavelengths, likely because of the presence of dust.  SN~2010jl has been suggested to have pre-existing dust \citep{2011AJ....142...45A}, post-shock dust formation \citep{2012AJ....143...17S,2013ApJ...776....5M,2014Natur.511..326G}, or no dust \citep{2012AJ....144..131Z,2014ApJ...797..118F}, but SN~2014ab does not show a more pronounced blueshifted component at shorter wavelengths, implying that the asymmetry we observe is not caused by small grain dust.

Spectropolarimetry of SN~2010jl indicates continuum polarization as high as $\sim 2$\% \citep{2011A&A...527L...6P,inprep}.  Additionally, depolarization of H$\alpha$ and H$\beta$ suggest that this continuum polarization reflects real asymmetries in the SN~2010jl photosphere.  Based on models with an axially symmetric prolate morphology, \citet{2015MNRAS.449.4304D} find a pole-to-equator density ratio of $\sim 2.6$.  On the other hand, SN~2014ab was observed to have very low continuum polarization at five epochs, and exhibits no sign of depolarization near H$\alpha$.

\begin{figure*}
\centering
\includegraphics[width=1\textwidth,height=1\textheight,keepaspectratio,clip=true,trim=0cm 0cm 0cm 0cm]{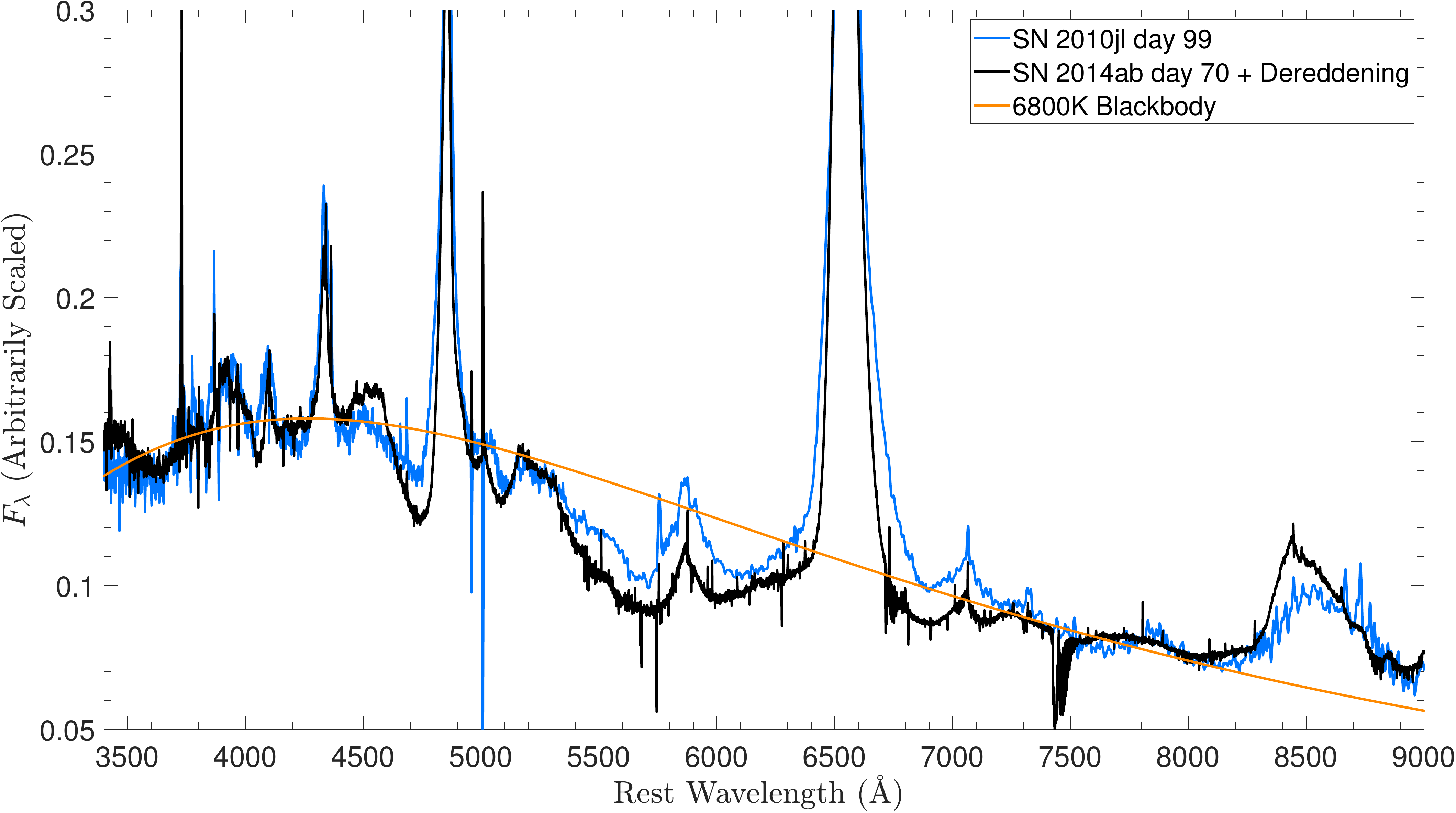}
\caption{A comparison of the spectra for SN~2014ab (day 70, VLT X-shooter) and SN~2010jl (Feb. 9, 2014, Lick/Kast, \citealt{2012AJ....143...17S}, corresponding to day 99) which have both been dereddened.  This spectrum of SN~2014ab was dereddened by an additional amount $E_{B-V}=0.10$\,mag beyond the  Milky Way and host-galaxy reddening applied earlier.  This additional amount was chosen so that the continuum slope of SN~2014ab would match the continuum shape of SN~2010jl for comparison of their spectral features.  We chose a SN~2010jl spectrum from a later relative date because the date of discovery for SN~2014ab was likely later relative to its peak.  A 6400\,K blackbody is shown in orange.  The optical spectral features are remarkably similar in SN~2014ab and SN~2010jl, though SN~2014ab clearly exhibits stronger broad P-Cygni absorption features and a stronger  Ca~\textsc{ii} near-IR triplet.}
\label{fig:LinSpec}
\end{figure*}

Another key difference between SN~2014ab and SN~2010jl is the prominence of their H$\beta$ absorption lines and Ca \textsc{ii} near-IR triplet emission lines, shown in Figure \ref{fig:LinSpec}.  SN~2014ab shows stronger absorption in H$\beta$ (as well as stronger broad absorption in H$\alpha$ and He~\textsc{i} $\lambda$5876 and $\lambda$7065) and stronger emission in the Ca \textsc{ii} near-IR triplet emission lines.  This suggests that the light from the SN photosphere that we are seeing from SN~2014ab passes through a larger amount of SN ejecta than that of SN~2010jl.

\subsection{Physical Picture for SN~2014ab}
\label{sec:Dis:PhysPic}
SN~2014ab is a peculiar, luminous SN~IIn.  It is puzzling in the sense that it does not show strong evidence for asymmetry in its low level of polarization, but asymmetry is evident from the line profiles in the spectra.  The large number of similarities between SN~2014ab and SN~2010jl provide motivation to consider a similar physical scenario for the two events, but certain key differences (polarization, broad P-Cygni absorption features) suggest that we may be looking at analogous events from different viewing angles.  

\begin{figure*}
\centering
\includegraphics[width=1\textwidth,height=1\textheight,keepaspectratio,clip=true,trim=0cm 0cm 0cm 0cm]{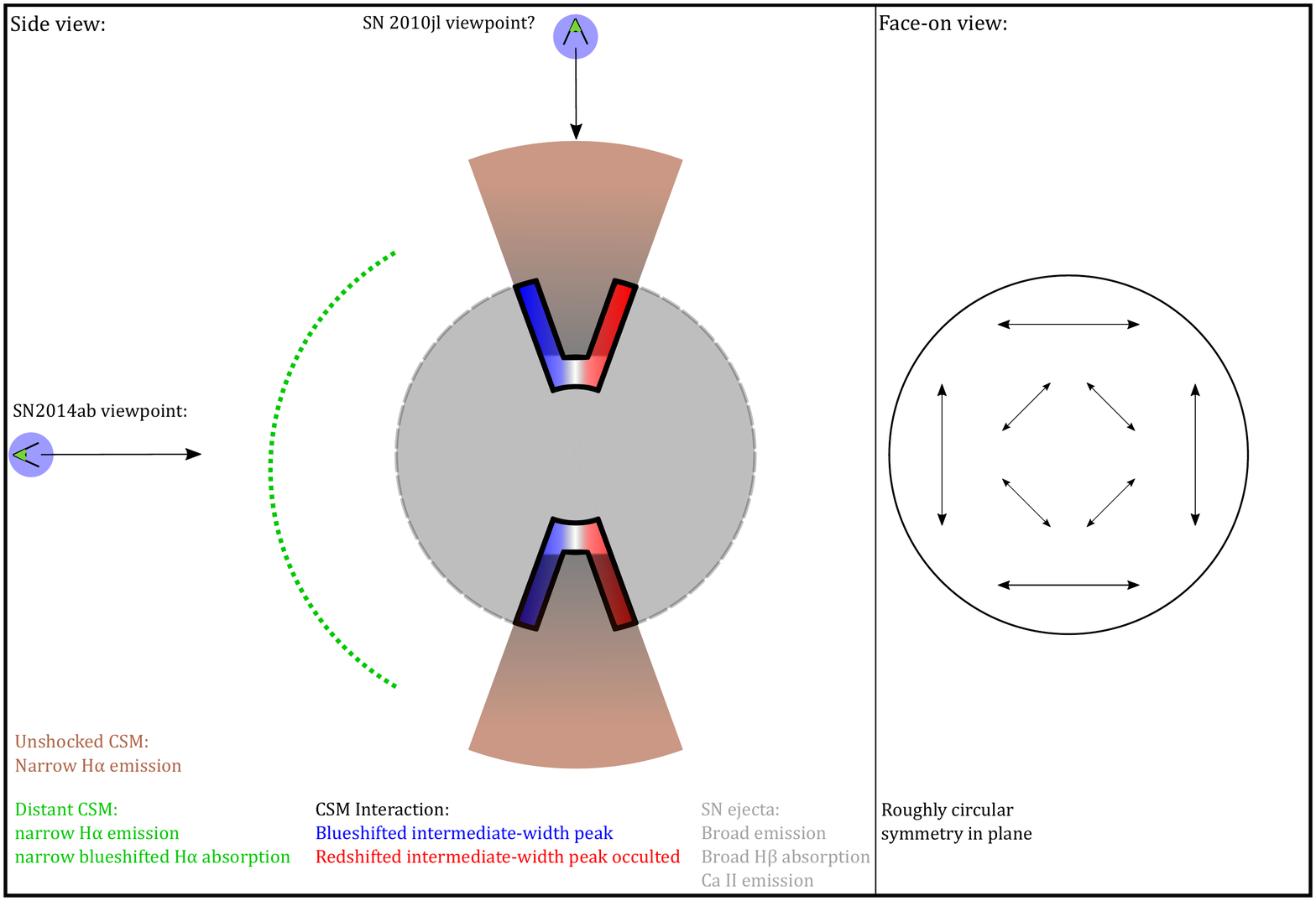}
\caption{A schematic showing how the viewing angle may significantly impact what line features and polarization are observed in SNe~IIn.  The prominent Ca~\textsc{ii} near-IR triplet emission, broad H$\beta$ absorption, and broad H$\alpha$ probably originate in the SN ejecta shaded in grey.  The CSM interaction region outlined in black is likely the source of the asymmetric blueshifted intermediate-width peak of the H$\alpha$ line profile, because the continuum photosphere in this zone occults the redshifted side.  The distant CSM shown in green causes the narrow blueshifted H$\alpha$ and H$\beta$ absorption.  Both the distant CSM in green and brown cause the narrow H$\alpha$ emission.  These features suggest that the likely viewing angle for SN~2014ab is from above the pole or axis of symmetry (from the side as labeled above).  For SN~2010jl, the observer may be seeing a similar geometry from a vantage point near the equatorial plane (from above in the schematic shown here).  The spectropolarimetric data suggest we see a roughly circularly symmetric SN from a perpendicular viewpoint.}
\label{fig:SN2014abCartoon}
\end{figure*}

The key features of SN~2014ab are interpreted in Figure \ref{fig:SN2014abCartoon}.  We suggest that the majority of its luminosity likely arises in an equatorially concentrated or bipolar CSM interaction region, which wraps around the densest equatorial CSM.  The continuum photosphere resides in this CSM interaction region and may therefore create a stronger blueshifted intermediate-width emission component if the continuum photosphere occults the redshifted side of the equatorial CSM interaction region.  This CSM interaction region is likely formed from mass lost from the progenitor of SN~2014ab within the 10--20\,yr prior to its death.  The SN ejecta from the explosion itself are likely optically thin at the time of our observations and have faded, but their presence can be seen in the broad components of hydrogen Balmer emission lines, Ca~\textsc{ii} near-IR triplet emission lines, and broad H$\beta$ absorption.

The narrow blueshifted H$\alpha$ and H$\beta$ absorption as well as the narrow H$\alpha$ emission persist throughout the evolution, so they likely arise from distant CSM along our line of sight that has not yet been overrun.  These features place the likely viewing angle for SN~2014ab along the axis of symmetry (i.e., above the pole), as labeled in Figure \ref{fig:SN2014abCartoon}, whereas SN~2010jl is probably observed from a different viewpoint shown in the schematic.  From the viewpoint of SN~2014ab, spectropolarimetric data could imply a roughly circularly symmetric photosphere because we are seeing a disk or torus face-on, whereas from the viewpoint of SN~2010jl, a significant polarization could be observed because we are seeing the same geometry edge-on.  \citet{2011AJ....142...45A} find a similar geometry with a torus inclined at 60$\degree$--80$\degree$ for SN~2010jl, while \citet{2017PhDT........74H} finds an inclination angle of 51$\degree$--74$\degree$.  \citet{2015MNRAS.449.4304D} also favour an edge-on view of SN~2010jl.

It is plausible that even if all SNe~IIn are significantly asymmetric in either their explosion or CSM interaction, a number of SNe~IIn would still exhibit unpolarized spectropolarimetric features simply because of a viewing-angle effect.  To date, no published studies of SNe~IIn examined with spectropolarimetric data have shown little to no polarization (i.e., less than 1\%) until this study of SN~2014ab.  It is therefore important to examine a larger population of SNe~IIn with spectropolarimetric data to determine if a similar geometry but with different orientations could account for the distinctions seen in SN~2014ab compared to other more typical SNe~IIn.   Efforts to model the polarization of H$\alpha$ profiles of SNe~IIn to better understand the importance of viewing angle \citep{2017IAUS..329..408H} should also be continued and expanded upon.

\section{Summary}

SN~2014ab is unusual compared to most Type IIn SNe with published spectropolarimetry in that it exhibits very low detectable continuum polarization ($0.07\%\pm 0.04\%$ to $0.43\%\pm 0.09\%$, with an upper limit of $<1.18$\%, limited primarily by the uncertainty in the interstellar polarization) in all five epochs of spectropolarimetry, whereas all other SNe~IIn show $>1$\% continuum polarization.  Although this lack of polarization implies near circular symmetry in the plane of the sky, the spectra of SN~2014ab display evidence for significant asymmetry along our line of sight based on radial veocities.  Without spectropolarimetric data, we would have simply concluded that SN~2014ab was a typical asymmetric SN~IIn without a constraint on our viewing angle.  Thus, it is crucial that spectropolarimetric data, even if it results in very low polarization measurements, is obtained in order to better constrain physical parameters of SNe~IIn. Below we summarise our main results.

\begin{itemize}
\item{SN~2014ab was already on the decline when first detected, and with the last predetection upper limit 212 days before detection, the date of explosion for SN~2014ab is uncertain. Its observed peak was $M_V =-19.14$\,mag at the time of first detection, but the true peak was likely somewhat brighter.  It declined very slowly, maintaining a brightness comparable to that of superluminous SNe for over 200 days.}
\item{Our spectra cover roughly 150 days and we see little evolution in the spectral features during this time.  This reinforces the conjecture that SN~2014ab was not first observed near its explosion date, but rather many weeks or months after explosion.}
\item{Narrow emission and absorption components of lines like H$\alpha$ and H$\beta$ indicate an expansion speed of 80\,km\,s$^{-1}$ for the pre-shock CSM. This likely corresponds to the progenitor wind speed or expansion speed of a pre-SN eruption.}
\item{Based on the sustained luminosity of SN~2014ab, the intermediate-width component of its H$\alpha$ emission line from post-shock gas, and its CSM speed, we estimate the mass-loss rate of the SN~2014ab progenitor to be $\dot{M} \approx \,{\rm M}_{\odot}\,\mathrm{yr^{-1}} (v_{\rm w}/(80\,\mathrm{km}\,\mathrm{s^{-1}}))$.  From the roughly 200 day duration of the high-luminosity phase, combined with the relative speeds of the CSM and shock, we infer a total mass loss of at least $\sim 5\,{\rm M}_{\odot}$ in the decade or so before explosion.}
\item{We measure the continuum polarization of SN~2014ab at five different epochs spanning days 76 to 165 to be between $0.07\%\pm 0.04\%$ and $0.43\%\pm 0.09\%$.  The ISP dominates the uncertainty in the true polarization, providing an upper limit of 1.18\%.  This suggests that SN~2014ab is not significantly polarized, unlike other SNe~IIn with published spectropolarimetry to date showing 1--3\% continuum polarization.}
\item{Spectra of SN~2014ab indicate a number of interesting features.  In particular, the H$\alpha$ and Pa$\beta$ line profiles exhibit similar shape at all epochs, implying little or no small dust grain formation.  Additionally, strong P-Cygni absorption is seen in H$\beta$ extending out to $-18,000$\,km\,$\mathrm{s^{-1}}$, suggesting that we are observing the continuum photosphere through a significant portion of the rapidly expanding SN ejecta.}
\item{Overall, SN~2014ab shows many similarities to SN~2010jl (see Fig. \ref{fig:LinSpec}).  Most importantly, both SNe have asymmetric H$\alpha$ emission features with an intermediate-width component offset blueward of the narrow component.  In the case of SN~2010jl, this asymmetric line profile evolves over time and is probably caused by dust formation \citep{2012AJ....143...17S}, in contrast to SN~2014ab.}
\item{Importantly, SN~2014ab differs from SN~2010jl in a number of ways.  Spectropolarimetry of SN~2014ab suggests low continuum polarization.  Spectra of SN~2014ab show broad P-Cygni absorption features and Ca~\textsc{ii} near-IR triplet emission features stronger than those seen in SN~2010jl.}
\item{We suggest that SN~2014ab and SN~2010jl could have quite similar CSM geometry but viewed from different directions.  SN~2014ab may be viewed nearly along the axis of symmetry (i.e., from above the pole), yielding low polarization due to its symmetry in the plane of the sky, whereas SN~2010jl may be viewed near edge-on, revealing larger polarization levels.}
\end{itemize}

As our paper was in the final stages of preparation, a preprint appeared that presented an independent analysis of similar spectroscopic data on SN~2014ab  \citep{2020arXiv200610198M}.  We briefly note some similarities and differences in the two analyses here.  Those authors similarly find SN~2014ab to be a luminous and slowly declining Type IIn SN with a blueshifted intermediate-width component in H$\alpha$ comparable to that of SN~2010jl. While \citet{2020arXiv200610198M}  do not present polarization data (as we do), they utilise mid-IR observations (whereas we do not). Based on the mid-IR data, they suggest that SN~2014ab had pre-existing dust in the CSM, but did not have significant formation of dust in the cool dense shell as SN~2010jl did \citep{2012AJ....143...17S}.  We draw a similar conclusion about the lack of dust formation in SN~2014ab based on spectral line profiles.   However, \citet{2020arXiv200610198M} argue that the blueshifted intermediate-width component of H$\alpha$ is symmetric, adequately fit by a Lorentzian, and can be explained by acceleration of the unshocked CSM by SN radiation.  In contrast, we find that the intermediate-width line profile is not symmetric and cannot be fit by a Lorentzian, and requires either obscuration by large dust grains (because the effect is not strongly wavelength dependent), occultation by optically thick material, or a lack of symmetry between the far side and near side of the interaction region.  Other than these points, our study and theirs find similar results.

\section*{Acknowledgments}
We thank Andrew Bigley, James Bradley, Kiera Fuller, Carolina Gould, Melissa L. Graham, Kevin Hayakawa, Minkyu Kim, Jon C. Mauerhan, and Isaac Shivvers for their assistance with some of the observations and reductions.

This work was supported by NSF grant AST-1210599.  N.S. received additional support from NSF grants AST-1312221 and AST-1515559, and by a Scialog grant from the Research Corporation for Science Advancement.  D.A.H. and G.H. were supported by NSF grant AST-1313484.  Research by D.J.S. is supported by NSF grants AST-1821967, 1821987, 1813708, 1813466, 1908972, and by the Heising-Simons Foundation under grant \#2020-1864.  The work of A.V.F.'s supernova group at U.C. Berkeley has been generously supported by NSF grant AST-1211916, the TABASGO Foundation, the Christopher R. Redlich Fund, and the Miller Institute for Basic Research in Science (U.C. Berkeley).  D.C.L. acknowledges support from NSF grants AST-1009571 and AST-1210311, under which part of this research was carried out.  X-shooter observations collected at the European Southern Observatory under ESO programme 092.D-0645(A) and EFOSC2 observations collected at the European Southern Observatory under ESO programme 191.D-0935 were used in this work.  This research used astrometric solutions from Astrometry.net. We also made use of the NASA/IPAC Extragalactic Database (NED), which is operated by the Jet Propulsion Laboratory, California Institute of Technology, under contract with NASA.  This work makes use of observations from the Las Cumbres Observatory network.

We thank the staffs at the MMT, Bok, Lick, European Southern Observatories, and Las Cumbres Observatories for their assistance with the observations.  Observations using Steward Observatory facilities were obtained as part of the large observing program AZTEC: Arizona Transient Exploration and Characterization. Some observations reported here were obtained at the MMT Observatory, a joint facility of the University of Arizona and the Smithsonian Institution.  Research at Lick Observatory is partially supported by a generous gift from Google.

\section*{Data Availability}
The data underlying this article will be shared on reasonable request to the corresponding author.

\appendix

\bsp

\label{lastpage}


\begin{thebibliography}{99}

\bibitem[Alard \& Lupton(1998)]{1998ApJ...503..325A} Alard, C., \& Lupton, R.~H.\ 1998, \apj, 503, 325

\bibitem[Andrews et al.(2017)]{2017MNRAS.471.4047A} Andrews, J.~E., Smith, N., McCully, C., et al.\ 2017, Monthly Notices of the Royal Astronomical Society, 471, 4047


\bibitem[Abazajian et al.(2009)]{2009ApJS..182..543A} Abazajian, K.~N., Adelman-McCarthy, J.~K., Ag{\"u}eros, M.~A., et al.\ 2009, \apjs, 182, 543 

\bibitem[Anderson et al.(2014)]{2014ApJ...786...67A} Anderson, J.~P., Gonz{\'a}lez-Gait{\'a}n, S., Hamuy, M., et al.\ 2014, \apj, 786, 67 

\bibitem[Andrews et al.(2011)]{2011AJ....142...45A} Andrews, J.~E., Clayton, G.~C., Wesson, R., et al.\ 2011, \aj, 142, 45 

\bibitem[Arcavi et al.(2012)]{2012ApJ...756L..30A} Arcavi, I., Gal-Yam, A., Cenko, S.~B., et al.\ 2012, \apjl, 756, L30 

\bibitem[Arcavi et al.(2014)]{2014ApJ...793...38A} Arcavi, I., Gal-Yam, A., Sullivan, M., et al.\ 2014, \apj, 793, 38 

\bibitem[Arnett et al.(1989)]{1989ARA&A..27..629A} Arnett, W.~D., Bahcall, J.~N., Kirshner, R.~P., \& Woosley, S.~E.\ 1989, \araa, 27, 629

\bibitem[Bilinski et al.(2015)]{2015MNRAS.450..246B} Bilinski, C., Smith, N., Li, W., et al.\ 2015, \mnras, 450, 246 

\bibitem[Bilinski et al.(2018)]{2018MNRAS.475.1104B} Bilinski, C., Smith, N., Williams, G.~G., et al.\ 2018, \mnras, 475, 1104 

\bibitem[Borish et al.(2015)]{2015ApJ...801....7B} Borish, H.~J., Huang, C., Chevalier, R.~A., et al.\ 2015, \apj, 801, 7 

\bibitem[Brown et al.(2013)]{2013PASP..125.1031B} Brown, T.~M., Baliber, N., Bianco, F.~B., et al.\ 2013, \pasp, 125, 1031

\bibitem[Burrows et al.(1995)]{1995ApJ...450..830B} Burrows, A., Hayes, J., \& Fryxell, B.~A.\ 1995, \apj, 450, 830   

\bibitem[Buzzoni et al.(1984)]{1984Msngr..38....9B} Buzzoni, B., Delabre, B., Dekker, H., et al.\ 1984, The Messenger, 38, 9 

\bibitem[Chevalier \& Fransson(1994)]{1994ApJ...420..268C} Chevalier, R.~A., \& Fransson, C.\ 1994, \apj, 420, 268 

\bibitem[Chevalier(2012)]{2012ApJ...752L...2C} Chevalier, R.~A.\ 2012, \apjl, 752, L2

\bibitem[Chornock et al.(2010)]{2010ApJ...713.1363C} Chornock, R., Filippenko, A.~V., Li, W., \& Silverman, J.~M.\ 2010, \apj, 713, 1363 

\bibitem[Chugai(2001)]{2001MNRAS.326.1448C} Chugai, N.~N.\ 2001, \mnras, 326, 1448 

\bibitem[Chugai(2006)]{2006AstL...32..739C} Chugai, N.~N.\ 2006, Astronomy Letters, 32, 739 

\bibitem[Chugai \& Danziger(1994)]{1994MNRAS.268..173C} Chugai, N.~N., \& Danziger, I.~J.\ 1994, \mnras, 268, 173   

\bibitem[Chugai et al.(2005)]{2005AstL...31..792C} Chugai, N.~N., Fabrika, S.~N., Sholukhova, O.~N., et al.\ 2005, Astronomy Letters, 31, 792 

\bibitem[Corcoran et al.(2001)]{2001ApJ...547.1034C} Corcoran, M.~F., Ishibashi, K., Swank, J.~H., \& Petre, R.\ 2001, \apj, 547, 1034 

\bibitem[Damineli(1996)]{1996ApJ...460L..49D} Damineli, A.\ 1996, \apjl, 460, L49 

\bibitem[Dessart et al.(2009)]{2009MNRAS.394...21D} Dessart, L., Hillier, D.~J., Gezari, S., Basa, S., \& Matheson, T.\ 2009, \mnras, 394, 21 

\bibitem[Dessart \& Hillier(2011)]{2011MNRAS.410.1739D} Dessart, L., \& Hillier, D.~J.\ 2011, \mnras, 410, 1739 

\bibitem[Dessart et al.(2015)]{2015MNRAS.449.4304D} Dessart, L., Audit, E., \& Hillier, D.~J.\ 2015, \mnras, 449, 4304   

\bibitem[Drake et al.(2009)]{2009ApJ...696..870D} Drake, A.~J., Djorgovski, S.~G., Mahabal, A., et al.\ 2009, \apj, 696, 870

\bibitem[Duquennoy \& Mayor(1991)]{1991A&A...248..485D} Duquennoy, A., \& Mayor, M.\ 1991, \aap, 248, 485 

\bibitem[Evans \& Kochanek(1989)]{1989ApJ...346L..13E} Evans, C.~R., \& Kochanek, C.~S.\ 1989, \apjl, 346, L13 

\bibitem[Faber et al.(2003)]{2003SPIE.4841.1657F} Faber, S.~M., Phillips, A.~C., Kibrick, R.~I., et al.\ 2003, \procspie, 4841, 1657   

\bibitem[Faran et al.(2014a)]{2014MNRAS.445..554F} Faran, T., Poznanski, D., Filippenko, A.~V., et al.\ 2014a, \mnras, 445, 554   

\bibitem[Faran et al.(2014b)]{2014MNRAS.442..844F} Faran, T., Poznanski, D., Filippenko, A.~V., et al.\ 2014b, \mnras, 442, 844   

\bibitem[Fassia et al.(2000)]{2000MNRAS.318.1093F} Fassia, A., Meikle, W.~P.~S., Vacca, W.~D., et al.\ 2000, \mnras, 318, 1093   

\bibitem[Filippenko(1982)]{1982PASP...94..715F} Filippenko, A.~V.\ 1982, \pasp, 94, 715 

\bibitem[Filippenko(1997)]{1997ARA&A} Filippenko, A.~V.\ 1997, \araa, 35, 309 

\bibitem[France et al.(2011)]{2011ApJ...743..186F} France, K., McCray, R., Penton, S.~V., et al.\ 2011, \apj, 743, 186 

\bibitem[Fransson et al.(2005)]{2005ApJ...622..991F} Fransson, C., Challis, P.~M., Chevalier, R.~A., et al.\ 2005, \apj, 622, 991   

\bibitem[Fransson et al.(2002)]{2002ApJ...572..350F} Fransson, C., Chevalier, R.~A., Filippenko, A.~V., et al.\ 2002, \apj, 572, 350   

\bibitem[Fransson et al.(2014)]{2014ApJ...797..118F} Fransson, C., Ergon, M., Challis, P.~J., et al.\ 2014, \apj, 797, 118   

\bibitem[Fransson et al.(2013)]{2013ApJ...768...88F} Fransson, C., Larsson, J., Spyromilio, J., et al.\ 2013, \apj, 768, 88 

\bibitem[Fraser et al.(2014)]{2014ATel.5968....1F} Fraser, M., Blagorodnova, N., Walton, N., et al.\ 2014, The Astronomer's Telegram, 5968,  

\bibitem[Gall et al.(2014)]{2014Natur.511..326G} Gall, C., Hjorth, J., Watson, D., et al.\ 2014, \nat, 511, 326 

\bibitem[Gal-Yam(2012)]{2012Sci...337..927G} Gal-Yam, A.\ 2012, Science, 337, 927 

\bibitem[Ganeshalingam et al.(2010)]{2010ApJS..190..418G} Ganeshalingam, M., Li, W., Filippenko, A.~V., et al.\ 2010, \apjs, 190, 418

\bibitem[Gezari et al.(2012)]{2012Natur.485..217G} Gezari, S., Chornock, R., Rest, A., et al.\ 2012, \nat, 485, 217 

\bibitem[Hill et al.(1998)]{LRS-HET}Hill, G.J., Nicklas, H.E., MacQueen, P.J., Tejada, C., Cobos Duenas, F.J., and Mitsch, W. 1998, Proc. SPIE, 3355, 375

\bibitem[Hillier et al.(2001)]{2001ApJ...553..837H} Hillier, D.~J., Davidson, K., Ishibashi, K., \& Gull, T.\ 2001, \apj, 553, 837 

\bibitem[Hoffman et al.(2008)]{2008ApJ...688.1186H} Hoffman, J.~L., Leonard, D.~C., Chornock, R., et al.\ 2008, \apj, 688, 1186-1209 

\bibitem[H\"{o}flich(1991)]{1991A&A...246..481H} H\"{o}flich, P.\ 1991, \aap, 246, 481 

\bibitem[Howerton et al.(2014)]{2014CBET.3826....1H} Howerton, S., Drake, A.~J., Djorgovski, S.~G., et al.\ 2014, Central Bureau Electronic Telegrams, 3826, 1 

\bibitem[Huk(2017)]{2017PhDT........74H} Huk, L.~N.\ 2017, Ph.D. Thesis

\bibitem[Huk(2017)]{2017IAUS..329..408H} Huk, L.\ 2017, The Lives and Death-throes of Massive Stars, 408

\bibitem[Kashi et al.(2013)]{2013MNRAS.436.2484K} Kashi, A., Soker, N., \& Moskovitz, N.\ 2013, \mnras, 436, 2484 

\bibitem[Katsuda et al.(2014)]{2014ApJ...780..184K} Katsuda, S., Maeda, K., Nozawa, T., Pooley, D., \& Immler, S.\ 2014, \apj, 780, 184   

\bibitem[Khokhlov et al.(1999)]{1999ApJ...524L.107K} Khokhlov, A.~M., H{\"o}flich, P.~A., Oran, E.~S., et al.\ 1999, \apjl, 524, L107 

\bibitem[Kiminki et al.(2016)]{2016MNRAS.463..845K} Kiminki, M.~M., Reiter, M., \& Smith, N.\ 2016, \mnras, 463, 845 

\bibitem[Lang et al.(2010)]{2010AJ....139.1782L} Lang, D., Hogg, D.~W., Mierle, K., et al.\ 2010, \aj, 139, 1782

\bibitem[Leonard \& Filippenko(2001)]{2001PASP..113..920L} Leonard, D.~C., \& Filippenko, A.~V.\ 2001, \pasp, 113, 920 

\bibitem[Leonard \& Filippenko(2005)]{2005ASPC..342..330L} Leonard, D.~C., \& Filippenko, A.~V.\ 2005, 1604-2004: Supernovae as Cosmological Lighthouses, 342, 330   

\bibitem[Leonard et al.(2001)]{2001ApJ...553..861L} Leonard, D.~C., Filippenko, A.~V., Ardila, D.~R., \& Brotherton, M.~S.\ 2001, \apj, 553, 861 

\bibitem[Leonard et al.(2000)]{2000ApJ...536..239L} Leonard, D.~C., Filippenko, A.~V., Barth, A.~J., \& Matheson, T.\ 2000, \apj, 536, 239   

\bibitem[Leonard et al.(2002)]{2002PASP..114...35L} Leonard, D.~C., Filippenko, A.~V., Gates, E.~L., et al.\ 2002, \pasp, 114, 35   

\bibitem[Leonard et al.(2002)]{2002AJ....124.2490L} Leonard, D.~C., Filippenko, A.~V., Li, W., et al.\ 2002, \aj, 124, 2490 

\bibitem[Leonard et al.(2006)]{2006Natur.440..505L} Leonard, D.~C., Filippenko, A.~V., Ganeshalingam, M., et al.\ 2006, \nat, 440, 505   

\bibitem[Leonard et al.(2012)]{2012AIPC.1429..204L} Leonard, D.~C., Dessart, L., Hillier, D.~J., \& Pignata, G.\ 2012, American Institute of Physics Conference Series, 1429, 204 

\bibitem[Leonard et al.(2016)]{2016IAUFM..29B.458L} Leonard, D.~C., Dessart, L., Pignata, G., et al.\ 2016, IAU Focus Meeting, 29, 458 

\bibitem[Lodato \& Rossi(2011)]{2011MNRAS.410..359L} Lodato, G., \& Rossi, E.~M.\ 2011, \mnras, 410, 359 

\bibitem[Maeda et al.(2013)]{2013ApJ...776....5M} Maeda, K., Nozawa, T., Sahu, D.~K., et al.\ 2013, \apj, 776, 5 

\bibitem[Marchenko et al.(2003)]{2003ApJ...596.1295M} Marchenko, S.~V., Moffat, A.~F.~J., Ballereau, D., et al.\ 2003, \apj, 596, 1295 

\bibitem[Mauerhan \& Smith(2012)]{2012MNRAS.424.2659M} Mauerhan, J., \& Smith, N.\ 2012, \mnras, 424, 2659 

\bibitem[Mauerhan et al.(2013)]{2013MNRAS.430.1801M} Mauerhan, J.~C., Smith, N., Filippenko, A.~V., et al.\ 2013, \mnras, 430, 1801 

\bibitem[Mauerhan et al.(2014)]{2014MNRAS.442.1166M} Mauerhan, J., Williams, G.~G., Smith, N., et al.\ 2014, \mnras, 442, 1166 

\bibitem[Mauerhan et al.(2017)]{2017ApJ...834..118M} Mauerhan, J.~C., Van Dyk, S.~D., Johansson, J., et al.\ 2017, \apj, 834, 118 

\bibitem[Maund et al.(2009)]{2009ApJ...705.1139M} Maund, J.~R., Wheeler, J.~C., Baade, D., et al.\ 2009, \apj, 705, 1139 

\bibitem[Miller \& Stone (1993)]{1993MillerStone} Miller, J. S., \& Stone, R. P. S. 1993, Lick Obs. Tech. Rep. 66 (Santa Cruz: Lick  Obs.)  

\bibitem[Miller et al. (1988)]{1988MillerSPOL} Miller, J. S., Robinson, L. B., \& Goodrich, R. W. 1988, Instrumentation for Ground-Based Optical Astronomy, 157  

\bibitem[Millour et al.(2009)]{2009A&A...506L..49M} Millour, F., Driebe, T., Chesneau, O., et al.\ 2009, \aap, 506, L49 

\bibitem[Moffat(1969)]{1969A&A.....3..455M} Moffat, A.~F.~J.\ 1969, \aap, 3, 455

\bibitem[Monnier et al.(1999)]{1999ApJ...525L..97M} Monnier, J.~D., Tuthill, P.~G., \& Danchi, W.~C.\ 1999, \apjl, 525, L97 

\bibitem[Monnier et al.(2002)]{2002ApJ...567L.137M} Monnier, J.~D., Tuthill, P.~G., \& Danchi, W.~C.\ 2002, \apjl, 567, L137 

\bibitem[Moriya et al.(2020)]{2020arXiv200610198M} Moriya, T.~J., Stritzinger, M.~D., Taddia, F., et al.\ 2020, arXiv e-prints, arXiv:2006.10198

\bibitem[Morozova et al.(2016)]{2016arXiv161008054M} Morozova, V., Piro, A.~L., \& Valenti, S.\ 2016, arXiv:1610.08054 

\bibitem[Munari \& Zwitter(1997)]{1997A&A...318..269M} Munari, U., \& Zwitter, T.\ 1997, \aap, 318, 269   

\bibitem[Nicholl et al.(2015)]{2015MNRAS.452.3869N} Nicholl, M., Smartt, S.~J., Jerkstrand, A., et al.\ 2015, \mnras, 452, 3869 

\bibitem[Ofek et al.(2014)]{2014ApJ...781...42O} Ofek, E.~O., Zoglauer, A., Boggs, S.~E., et al.\ 2014, \apj, 781, 42

\bibitem[Oke(1990)]{1990AJ.....99.1621O} Oke, J.~B.\ 1990, \aj, 99,1621


\bibitem[Parkin et al.(2009)]{2009MNRAS.394.1758P} Parkin, E.~R., Pittard, J.~M., Corcoran, M.~F., Hamaguchi, K., \& Stevens, I.~R.\ 2009, \mnras, 394, 1758 

\bibitem[Pastorello et al.(2013)]{2013ApJ...767....1P} Pastorello, A., Cappellaro, E., Inserra, C., et al.\ 2013, \apj, 767, 1 

\bibitem[Patat et al.(2011)]{2011A&A...527L...6P} Patat, F., Taubenberger, S., Benetti, S., Pastorello, A., \& Harutyunyan, A.\ 2011, \aap, 527, LL6   

\bibitem[Pei(1992)]{1992ApJ...395..130P} Pei, Y.~C.\ 1992, \apj, 395, 130 

\bibitem[Peterson(2001)]{2001sac..conf....3P} Peterson, B.~M.\ 2001, Advanced Lectures on the Starburst-AGN, 3 

\bibitem[Phillips et al.(2013)]{2013ApJ...779...38P} Phillips, M.~M., Simon, J.~D., Morrell, N., et al.\ 2013, \apj, 779, 38 

\bibitem[Phinney(1989)]{1989IAUS..136..543P} Phinney, E.~S.\ 1989, The Center of the Galaxy, 136, 543 

\bibitem[Poon et al.(2011)]{2011arXiv1109.0899P} Poon, H., Pun, J.~C.~S., Lam, T.~Y., Qiu, Y.~L., \& Wei, J.~Y.\ 2011, arXiv:1109.0899 

\bibitem[Porter et al.(2016)]{2016ApJ...828...24P} Porter, A.~L., Leising, M.~D., Williams, G.~G., et al.\ 2016, \apj, 828, 24

\bibitem[Poznanski et al.(2009)]{2009ApJ...694.1067P} Poznanski, D., Butler, N., Filippenko, A.~V., et al.\ 2009, \apj, 694, 1067   

\bibitem[Poznanski et al.(2011)]{2011MNRAS.415L..81P} Poznanski, D., Ganeshalingam, M., Silverman, J.~M.,   \& Filippenko, A.~V.\ 2011, \mnras, 415, L81   

\bibitem[Poznanski et al.(2012)]{2012MNRAS.426.1465P} Poznanski, D., Prochaska, J.~X., \& Bloom, J.~S.\ 2012, \mnras, 426, 1465   

\bibitem[Prieto et al.(2013)]{2013ApJ...763L..27P} Prieto, J.~L., Brimacombe, J., Drake, A.~J., \& Howerton, S.\ 2013, \apjl, 763, L27 

\bibitem[Quataert \& Shiode(2012)]{2012MNRAS.423L..92Q} Quataert, E., \& Shiode, J.\ 2012, \mnras, 423, L92

\bibitem[Quimby(2006)]{2006PhDT........13Q} Quimby, R.~M.\ 2006, Ph.D.~Thesis,  
\bibitem[Rees(1988)]{1988Natur.333..523R} Rees, M.~J.\ 1988, \nat, 333, 523 

\bibitem[Reilly et al.(2017)]{2017arXiv170108885R} Reilly, E., Maund, J.~R., Baade, D., et al.\ 2017, arXiv:1701.08885 

\bibitem[Richmond et al.(1994)]{1994AJ....107.1022R} Richmond, M.~W., Treffers, R.~R., Filippenko, A.~V., et al.\ 1994, \aj, 107, 1022 

\bibitem[Riess et al.(2005)]{2005ApJ...627..579R} Riess, A.~G., Li, W., Stetson, P.~B., et al.\ 2005, \apj, 627, 579   

\bibitem[Ryder et al.(2004)]{2004MNRAS.349.1093R} Ryder, S.~D., Sadler, E.~M., Subrahmanyan, R., et al.\ 2004, \mnras, 349, 1093 

\bibitem[Sana et al.(2012)]{2012Sci...337..444S} Sana, H., de Mink, S.~E., de Koter, A., et al.\ 2012, Science, 337, 444 

\bibitem[Schlafly \& Finkbeiner(2011)]{2011ApJ...737..103S} Schlafly, E.~F., \& Finkbeiner, D.~P.\ 2011, \apj, 737, 103 

\bibitem[Schlegel(1990)]{1990MNRAS.244..269S} Schlegel, E.~M.\ 1990, \mnras, 244, 269  

\bibitem[Schlegel(1996)]{1996AJ....111.1660S} Schlegel, E.~M.\ 1996, \aj, 111, 1660   

\bibitem[Schmidt et al.(1992)]{1992AJ....104.1563S} Schmidt, G.~D., Elston, R., \& Lupie, O.~L.\ 1992, \aj, 104, 1563   

\bibitem[Schmidt et al.(1992)]{1992ApJ...398L..57S} Schmidt, G.~D., Stockman, H.~S., \& Smith, P.~S.\ 1992, \apjl, 398, L57   

\bibitem[Serkowski et al.(1975)]{1975ApJ...196..261S} Serkowski, K., Mathewson, D.~S., \& Ford, V.~L.\ 1975, \apj, 196, 261   

\bibitem[Shivvers et al.(2015)]{2015ApJ...806..213S} Shivvers, I., Groh, J.~H., Mauerhan, J.~C., et al.\ 2015, \apj, 806, 213   

\bibitem[Smartt(2009)]{2009ARA&A..47...63S} Smartt, S.~J.\ 2009, \araa, 47, 63 

\bibitem[Smith(2010)]{2010MNRAS.402..145S} Smith, N.\ 2010, \mnras, 402, 145

\bibitem[Smith(2011)]{2011MNRAS.415.2020S} Smith, N.\ 2011, \mnras, 415, 2020 

\bibitem[Smith(2014)]{2014ARA&A..52..487S} Smith, N.\ 2014, \araa, 52, 487 

\bibitem[Smith(2017)]{smith17} Smith N., 2017, in Alsabti A.W., Murdin P., eds, Handbook of Supernovae.
Springer International Publishing, Berlin, p. 403

\bibitem[Smith \& Arnett(2014)]{2014ApJ...785...82S} Smith, N., \& Arnett, W.~D.\ 2014, \apj, 785, 82 

\bibitem[Smith \& McCray(2007)]{sm07} Smith, N., \& McCray, R. 2007, ApJ, 671, L17

\bibitem[Smith \& Owocki(2006)]{so06} Smith, N., \& Owocki, S.P.\ 2006, ApJ, 645, L45

\bibitem[Smith, Ginsburg, \& Bally (2017)]{SmithALMAsubmitted} Smith, N., Ginsburg, A., \& Bally, J.\ 2017, submitted

\bibitem[Smith et al.(2007)]{smith07} Smith, N., et al. 2007, ApJ, 666, 1116

\bibitem[Smith et al.(2008)]{2008ApJ...686..467S} Smith, N., Chornock, R., Li, W., et al.\ 2008, \apj, 686, 467-484 

\bibitem[Smith et al.(2010)]{2010ApJ...709..856S} Smith, N., Chornock, R., Silverman, J.~M., Filippenko, A.~V., \& Foley, R.~J.\ 2010, \apj, 709, 856   

\bibitem[Smith et al.(2008)]{2008ApJ...680..568S} Smith, N., Foley, R.~J., \& Filippenko, A.~V.\ 2008, \apj, 680, 568

\bibitem[Smith et al.(2011)]{2011MNRAS.412.1522S} Smith, N., Li, W., Filippenko, A.~V., \& Chornock, R.\ 2011, \mnras, 412, 1522   

\bibitem[Smith et al.(2015)]{2015MNRAS.449.1876S} Smith, N., Mauerhan, J.~C., Cenko, S.~B., et al.\ 2015, \mnras, 449, 1876   

\bibitem[Smith et al.(2014)]{2014MNRAS.438.1191S} Smith, N., Mauerhan, J.~C., \& Prieto, J.~L.\ 2014, \mnras, 438, 1191 

\bibitem[Smith et al.(2009)]{2009ApJ...695.1334S} Smith, N., Silverman, J.~M., Chornock, R., et al.\ 2009, \apj, 695, 1334 

\bibitem[Smith et al.(2012)]{2012AJ....143...17S} Smith, N., Silverman, J.~M., Filippenko, A.~V., et al.\ 2012, \aj, 143, 17   

\bibitem[Smith \& Tombleson(2015)]{2015MNRAS.447..598S} Smith, N., \& Tombleson, R.\ 2015, \mnras, 447, 598 

\bibitem[Smith et al.(2005)]{2005ApJ...635L..41S} Smith, N., Zhekov, S.~A., Heng, K., et al.\ 2005, \apjl, 635, L41 	

\bibitem[Stahl et al.(2019)]{2019MNRAS.490.3882S} Stahl, B.~E., Zheng, W., de Jaeger, T., et al.\ 2019, \mnras, 490, 3882

\bibitem[Stetson(1987)]{1987PASP...99..191S} Stetson, P.~B.\ 1987, \pasp, 99, 191 

\bibitem[Stritzinger et al.(2012)]{2012ApJ...756..173S} Stritzinger, M., Taddia, F., Fransson, C., et al.\ 2012, \apj, 756, 173   

\bibitem[Strubbe \& Quataert(2009)]{2009MNRAS.400.2070S} Strubbe, L.~E., \& Quataert, E.\ 2009, \mnras, 400, 2070 

\bibitem[Sugawara et al.(2015)]{2015PASJ...67..121S} Sugawara, Y., Maeda, Y., Tsuboi, Y., et al.\ 2015, \pasj, 67, 121 

\bibitem[Turatto et al.(2003)]{2003fthp.conf..200T} Turatto, M., Benetti, S., \& Cappellaro, E.\ 2003, From Twilight to Highlight: The Physics of Supernovae, 200 

\bibitem[Tuthill et al.(1999)]{1999Natur.398..487T} Tuthill, P.~G., Monnier, J.~D., \& Danchi, W.~C.\ 1999, \nat, 398, 487 

\bibitem[Ulrich et al.(1997)]{1997ARA&A..35..445U} Ulrich, M.-H., Maraschi, L., \& Urry, C.~M.\ 1997, \araa, 35, 445 

\bibitem[Valenti et al.(2016)]{2016MNRAS.459.3939V} Valenti, S., Howell, D.~A., Stritzinger, M.~D., et al.\ 2016, \mnras, 459, 3939 

\bibitem[Vernet et al.(2011)]{2011A&A...536A.105V} Vernet, J., Dekker, H., D'Odorico, S., et al.\ 2011, \aap, 536, A105 

\bibitem[Vinko et al.(2012)]{2012CBET.3022....1V} Vinko, J., Zheng, W., Marion, G.~H., et al.\ 2012, Central Bureau Electronic Telegrams, 3022, 1  

\bibitem[Vorontsov-Velyaminov(1959)]{1959VV....C......0V} Vorontsov-Velyaminov, B.~A.\ 1959, Atlas and Catalog of Interacting Galaxies (1959, 0

\bibitem[Wang et al.(2001)]{2001ApJ...550.1030W} Wang, L., Howell, D.~A., H{\"o}flich, P., \& Wheeler, J.~C.\ 2001, \apj, 550, 1030 

\bibitem[Wang et al.(2002)]{2002ApJ...579..671W} Wang, L., Wheeler, J.~C., H{\"o}flich, P., et al.\ 2002, \apj, 579, 671 

\bibitem[Wang \& Wheeler(2008)]{2008ARA&A..46..433W} Wang, L., \& Wheeler, J.~C.\ 2008, \araa, 46, 433 

\bibitem[Wang et al.(1997)]{1997ApJ...476L..27W} Wang, L., Wheeler, J.~C., \& H{\"o}flich, P.\ 1997, \apjl, 476, L27

\bibitem[Wardle \& Kronberg(1974)]{1974ApJ...194..249W} Wardle, J.~F.~C., \& Kronberg, P.~P.\ 1974, \apj, 194, 249 

\bibitem[Wheeler et al.(2002)]{2002ApJ...568..807W} Wheeler, J.~C., Meier, D.~L., \& Wilson, J.~R.\ 2002, \apj, 568, 807 

\bibitem[Williams et. al.(2020, in prep.)]{inprep} G. G. Williams, J. L. Hoffman, N. Smith, et al. in preparation

\bibitem[Woosley(2017)]{2017ApJ...836..244W} Woosley, S.~E.\ 2017, \apj, 836, 244

\bibitem[Yuan \& Akerlof(2008)]{2008ApJ...677..808Y} Yuan, F., \& Akerlof, C.~W.\ 2008, \apj, 677, 808 

\bibitem[Zhang et al.(2012)]{2012AJ....144..131Z} Zhang, T., Wang, X., Wu, C., et al.\ 2012, \aj, 144, 131 

\bibitem[Zhekov et al.(2014a)]{2014MNRAS.445.1663Z} Zhekov, S.~A., Tomov, T., Gawronski, M.~P., et al.\ 2014, \mnras

\bibitem[Zhekov et al.(2014b)]{2014ApJ...785....8Z} Zhekov, S.~A., Gagn{\'e}, M., \& Skinner, S.~L.\ 2014, \apj, 785, 8 


\end{thebibliography}
\end{document}